%% file: baryon.tex
\documentclass[12pt]{article}
\usepackage{epsfig}
\usepackage{cite}
\usepackage{amsmath}
\usepackage{hhline}
\usepackage{amssymb}
\usepackage{times}

\newlength{\dinwidth}
\newlength{\dinmargin}
\begin{document}  
\newcommand{\slowpi}{\pi_{\mathit{slow}}}
\newcommand{\fiidiii}{F_2^{D(3)}}
\newcommand{\fiidiiiarg}{\fiidiii\,(\beta,\,Q^2,\,x)}
\newcommand{\n}{1.19\pm 0.06 (stat.) \pm0.07 (syst.)}
\newcommand{\nz}{1.30\pm 0.08 (stat.)^{+0.08}_{-0.14} (syst.)}
\newcommand{\fiidiiiful}{F_2^{D(4)}\,(\beta,\,Q^2,\,x,\,t)}
\newcommand{\fiipom}{\tilde F_2^D}
\newcommand{\ALPHA}{1.10\pm0.03 (stat.) \pm0.04 (syst.)}
\newcommand{\ALPHAZ}{1.15\pm0.04 (stat.)^{+0.04}_{-0.07} (syst.)}
\newcommand{\fiipomarg}{\fiipom\,(\beta,\,Q^2)}
\newcommand{\pomflux}{f_{\pom / p}}
\newcommand{\nxpom}{1.19\pm 0.06 (stat.) \pm0.07 (syst.)}
\newcommand {\gapprox}
   {\raisebox{-0.7ex}{$\stackrel {\textstyle>}{\sim}$}}
\newcommand {\lapprox}
   {\raisebox{-0.7ex}{$\stackrel {\textstyle<}{\sim}$}}
\def\gsim{\,\lower.25ex\hbox{$\scriptstyle\sim$}\kern-1.30ex%
\raise 0.55ex\hbox{$\scriptstyle >$}\,}
\def\lsim{\,\lower.25ex\hbox{$\scriptstyle\sim$}\kern-1.30ex%
\raise 0.55ex\hbox{$\scriptstyle <$}\,}
\newcommand{\pomfluxarg}{f_{\pom / p}\,(x_\pom)}
\newcommand{\dsf}{\mbox{$F_2^{D(3)}$}}
\newcommand{\dsfva}{\mbox{$F_2^{D(3)}(\beta,Q^2,x_{I\!\!P})$}}
\newcommand{\dsfvb}{\mbox{$F_2^{D(3)}(\beta,Q^2,x)$}}
\newcommand{\dsfpom}{$F_2^{I\!\!P}$}
\newcommand{\gap}{\stackrel{>}{\sim}}
\newcommand{\lap}{\stackrel{<}{\sim}}
\newcommand{\fem}{$F_2^{em}$}
\newcommand{\tsnmp}{$\tilde{\sigma}_{NC}(e^{\mp})$}
\newcommand{\tsnm}{$\tilde{\sigma}_{NC}(e^-)$}
\newcommand{\tsnp}{$\tilde{\sigma}_{NC}(e^+)$}
\newcommand{\st}{$\star$}
\newcommand{\sst}{$\star \star$}
\newcommand{\ssst}{$\star \star \star$}
\newcommand{\sssst}{$\star \star \star \star$}
\newcommand{\tw}{\theta_W}
\newcommand{\sw}{\sin{\theta_W}}
\newcommand{\cw}{\cos{\theta_W}}
\newcommand{\sww}{\sin^2{\theta_W}}
\newcommand{\cww}{\cos^2{\theta_W}}
\newcommand{\trm}{m_{\perp}}
\newcommand{\trp}{p_{\perp}}
\newcommand{\trmm}{m_{\perp}^2}
\newcommand{\trpp}{p_{\perp}^2}
\newcommand{\alp}{\alpha_s}

\newcommand{\alps}{\alpha_s}
\newcommand{\sqrts}{$\sqrt{s}$}
\newcommand{\LO}{$O(\alpha_s^0)$}
\newcommand{\Oa}{$O(\alpha_s)$}
\newcommand{\Oaa}{$O(\alpha_s^2)$}
\newcommand{\PT}{p_{\perp}}
\newcommand{\JPSI}{J/\psi}
\newcommand{\sh}{\hat{s}}
\newcommand{\uh}{\hat{u}}
\newcommand{\MP}{m_{J/\psi}}
\newcommand{\PO}{I\!\!P}
\newcommand{\xbj}{x}
\newcommand{\xpom}{x_{\PO}}
\newcommand{\ttbs}{\char'134}
\newcommand{\xpomlo}{3\times10^{-4}}  
\newcommand{\xpomup}{0.05}  
\newcommand{\dgr}{^\circ}
\newcommand{\pbarnt}{\,\mbox{{\rm pb$^{-1}$}}}
\newcommand{\gev}{\,\mbox{GeV}}
\newcommand{\WBoson}{\mbox{$W$}}
\newcommand{\fbarn}{\,\mbox{{\rm fb}}}
\newcommand{\fbarnt}{\,\mbox{{\rm fb$^{-1}$}}}
\newcommand{\pom}{{\rm I \hspace{-0.3ex} \rm P}}
\newcommand{\reg}{{\rm I \hspace{-0.3ex} \rm R}}
\newcommand{\um}{\rm{\mu m}}
\newcommand {\sub}[1]{_{\mathrm{#1}}}
\newcommand{\der}{{\mathrm d}}
%
%
\newcommand{\qsq}{\ensuremath{Q^2} }
\newcommand{\gevsq}{\ensuremath{\mathrm{GeV}^2} }
\newcommand{\et}{\ensuremath{E_t^*} }
\newcommand{\rap}{\ensuremath{\eta^*} }
\newcommand{\gp}{\ensuremath{\gamma^*}p }
\newcommand{\dsiget}{\ensuremath{{\rm d}\sigma_{ep}/{\rm d}E_t^*} }
\newcommand{\dsigrap}{\ensuremath{{\rm d}\sigma_{ep}/{\rm d}\eta^*} }
\def\Journal#1#2#3#4{{#1} {\bf #2} (#3) #4}
\def\NCA{\em Nuovo Cimento}
\def\NIM{\em Nucl. Instrum. Methods}
\def\NIMA{{\em Nucl. Instrum. Methods} {\bf A}}
\def\NPB{{\em Nucl. Phys.}   {\bf B}}
\def\PLB{{\em Phys. Lett.}   {\bf B}}
\def\PRL{\em Phys. Rev. Lett.}
\def\PRD{{\em Phys. Rev.}    {\bf D}}
\def\ZPC{{\em Z. Phys.}      {\bf C}}
\def\EJC{{\em Eur. Phys. J.} {\bf C}}
\def\CPC{\em Comp. Phys. Commun.}

\begin{titlepage}

\noindent
DESY 98-169  \hspace{4.0in}  ISSN 0418--9833 \\
October 1998 \\

\vspace{2cm}

\begin{center}
\begin{Large}

{\bf \boldmath Measurement of Leading Proton and Neutron Production \\
   in Deep Inelastic Scattering at HERA}

\vspace{2cm}

H1 Collaboration

\end{Large}
\end{center}

\vspace{2cm}

\begin{abstract}
\noindent
Deep--inelastic scattering events with a leading baryon have been detected 
by the H1 experiment at HERA using a forward proton spectrometer and a 
forward neutron calorimeter. Semi--inclusive cross sections have been 
measured in the kinematic region
$2 \le Q^2 \le 50$~{\rm GeV}$^2$, $6\cdot 10^{-5} \le x \le 6\cdot 10^{-3}$ 
and baryon $p_T \le 200$~{\rm MeV}, for events with a final state proton with 
energy $580 \le E^{\prime} \le 740$~{\rm GeV}, or a neutron with energy 
$E^{\prime} \ge 160$~{\rm GeV}. The measurements are used to test production 
models and factorization hypotheses. A Regge model of leading baryon 
production which consists of pion, pomeron and secondary reggeon exchanges 
gives an acceptable description of both semi--inclusive cross sections 
in the region $0.7 \le E^{\prime} / E_p \le 0.9$, where $E_p$ is the 
proton beam energy. The leading neutron data are used to estimate for 
the first time the structure function of the pion at small Bjorken--$x$.
\end{abstract}

\vspace{1.5cm}

\begin{center}
Submitted to {\it Eur. Phys. J.}
\end{center}

\end{titlepage}

\newpage

\begin{sloppypar}
\input{h1auts}
\bigskip

\noindent
{\footnotesize{\input{h1inst}}}
\end{sloppypar}

\newpage

\section{Introduction}

\noindent
We report the measurement of the semi--inclusive cross sections for 
proton and neutron production in deep--inelastic scattering (DIS).
The $e p \rightarrow e N X$ data, where $N$ represents either 
a final state proton or neutron, were obtained during 1995 and 1996 
using the HERA accelerator at DESY where 27.5~{\rm GeV} positrons collided 
with 820~{\rm GeV} protons. Events containing leading baryons were 
detected using the H1 detector upgraded with a forward proton spectrometer 
(FPS) and a forward neutron calorimeter (FNC). 

The two multi--purpose experiments at HERA, H1 and ZEUS, have observed 
a class of events which are characterized by the absence of final state 
particles in the region of phase space between the outgoing proton debris 
and the current jet~\cite{ZEUSDIFF,H1DIFF}. These so called rapidity gap 
events contribute approximately 10\% to the total DIS cross section
and can be interpreted as being mainly due to interactions 
of a virtual photon with a pomeron~\cite{H1F2D3,ZEUSF2D3,H1F2D32,ZEUSF2D32}. 
In addition to virtual 
photon--pomeron interactions, one also expects meson exchanges to contribute 
to the total DIS cross section and to the production of leading protons and 
neutrons with small $p_T$~\cite{Sullivan,Holtmann,Povh,Szczurek}. Due to its 
small mass the most obvious candidate for such an exchange is the pion. 
If only pion exchange is responsible for leading baryon production and
if isospin is conserved at the proton vertex, the 
ratio of neutron and proton production should be equal to two due to the 
difference in the Clebsch--Gordon coefficients for the
$\pi^+ n$ and $\pi^0 p$ isospin $\frac{1}{2}$ states.

In this paper we present measurements of the semi--inclusive cross
sections for proton and neutron production in the same kinematic region 
of $x$, $Q^2$ and $p_T$, where $x$ is the Bjorken scaling variable,
$Q^2$ is the negative four--momentum squared of the virtual photon and
$p_T$ is the transverse momentum of the final state baryon with respect
to the beam direction. The measurements are compared to the results of a 
Regge model of leading baryon production and are used to test the pion
exchange expectation for the ratio of neutron and proton production. 
We also compare our measurements to the predictions of the 
LEPTO~6.5 Monte Carlo program~\cite{Lepto65} which simulates baryon
production without Regge dynamics by using an alternative formalism
based upon soft colour interactions~\cite{Buch,Lepto-soft} and the 
string fragmentation model~\cite{Jetset}. Our cross section measurements 
are relevant for the determination of fracture functions which are a 
perturbative QCD approach for describing the semi--inclusive production 
of hadrons in the proton fragmentation 
region~\cite{Trentadue,Florian,Grazzini}.

We use the kinematic variables $x$, $Q^2$ and $y$ to describe the inclusive 
DIS scattering process. They are defined as:
\begin{equation}
x=\frac{-q^2}{2 p \cdot q}
\hspace{4em}
Q^2=-q^2
\hspace{4em}
y=\frac{p \cdot q}{p \cdot k}\;\;,
\end{equation}
where $p$, $k$ and $q$ are the four--momenta of the incident proton, the
incident positron and the exchanged vector boson $(\gamma^*)$ coupling
to the positron. At $e p$ centre--of--mass energy $\sqrt{s}$ they are related
by $Q^2 = s x y$. 

The kinematic variables used to describe a final state baryon are:
\begin{equation}
t = (p-p^\prime)^2  \simeq  - \frac{p_T^2}{z} - (1-z) 
    {\Bigg (} \frac{m_N^2}{z} - m_p^2 {\Bigg )}
\hspace{2em}  z = 1-\frac{q \cdot (p-p^\prime)}{q \cdot p} 
    \simeq E^{\prime}/E_p, 
\end{equation}
where $p^\prime$ is the four--momentum of the final state baryon,
$m_N$ is the mass of the final state baryon and $m_p$ is the proton mass.
As defined, $t$ corresponds to the squared four--momentum transferred
between the incident proton and the final state baryon. It is
different from the ``rapidity gap'' definition $t = (p-p_Y)^2$
(used for example in reference~\cite{H1F2D32}), which defines the
hadronic system $Y$ at the proton vertex by the presence of a 
rapidity gap.
For leading baryon production the definition of $t$ in this paper
is more appropriate even though it is only equal to the 
rapidity gap definition for events where the final state baryon 
is exclusively produced at the proton vertex.

The four--fold differential cross section for baryon production
can be parameterized by a semi--inclusive structure function, 
${\mathrm{F_2^{LB(4)}}}$, defined by: 
\begin{eqnarray}
  \frac{\der^4 \sigma(e p \rightarrow e N X)}
       {\der x\,\der Q^2\,\der z\, \der t} = 
   \frac {4 \pi \alpha^2}{x\,Q^4} 
   \left ( 1 - y + \frac{y^2}{2 [1 + R(x,Q^2,z,t)]} \right )
   {\mathrm{F_2^{LB(4)}}}(x,Q^2,z,t),
\end{eqnarray}
where $\alpha$ is the fine structure constant and $R$ is the ratio 
between the absorption cross sections for longitudinally and transversely 
polarized virtual photons. In the kinematic range covered by this analysis,
the structure function ${\mathrm{F_2^{LB(4)}}}$ is rather insensitive to the
value of $R$ and we assume that $R = 0$. The variation from $R = 0$ to
$ R = \infty$ leads to a $9$\% change at most in the resulting value of 
${\mathrm{F_2^{LB(4)}}}$ for the range of $y$ covered in this measurement.

The four--fold differential cross section integrated over 
$0 \le p_T \le 200$~{\rm MeV} defines the semi--inclusive structure function
${\mathrm{F_2^{LB(3)}}}$ which we measure:
\begin{eqnarray}
  \frac{\der^3 \sigma (e p \rightarrow e N X)}
       {\der x\,\der Q^2\,\der z} & = & \int_{t_0}^{t_{{\mathrm{min}}}}
  \frac {4 \pi \alpha^2}{x\,Q^4} 
  \left ( 1 - y + \frac{y^2}{2} \right )
   {\mathrm{F_2^{LB(4)}}}(x,Q^2,z,t) \, \der t \\
   & = &
  \frac {4 \pi \alpha^2}{x\,Q^4} 
  \left ( 1 - y + \frac{y^2}{2} \right )
   {\mathrm{F_2^{LB(3)}}}(x,Q^2,z), \nonumber
\end{eqnarray}
where the integration limits are:
\begin{equation}
\label{tlimits}
t_{{\mathrm{min}}}  = - (1-z) {\Bigg (} \frac{m_N^2}{z} - m_p^2 {\Bigg )} 
\hspace{3em}
t_0 =  -\frac{(200 \, {\mathrm{MeV}})^2}{z}+t_{{\mathrm{min}}}.
\end{equation}
The structure function ${\mathrm{F_2^{LB(3)}}}$ is 
denoted by ${\mathrm{F_2^{LP(3)}}}$ and ${\mathrm{F_2^{LN(3)}}}$ 
for the  semi--inclusive processes which have final state protons and 
neutrons respectively.
We present measurements of ${\mathrm{F_2^{LP(3)}}}$
in the range {\mbox{$0.73 \le z \le 0.88$}} and of ${\mathrm{F_2^{LN(3)}}}$ 
for {\mbox{$0.3 \le z \le 0.9$}}. 
In this paper we do not discuss the very high $z$ region $(z \gsim 0.95)$
which is most relevant for pomeron exchange and which has been used
to measure the diffractive structure function 
${\mathrm{F_2^{D(3)}}}$~\cite{H1F2D3,ZEUSF2D3,H1F2D32,ZEUSF2D32}.

\section{H1 Apparatus}

\noindent
The central H1 detector is described in detail in 
references~\cite{H1detector,SPACALE1,SpaCal}.
Here only the forward proton spectrometer, the forward
neutron calorimeter and the crucial parts of the central H1 detector
used in this analysis are described. The coordinate system
convention used by the H1 experiment defines the positive $z$--axis,
or ``forward'' direction, as being that of the outgoing proton beam.
The positive $x$--axis points towards the centre of the HERA ring.

The components of the central H1 detector which are essential for this
analysis are the backward electromagnetic calorimeter (SPACAL),
the liquid argon calorimeter (LAr) and the central and forward
tracking systems. The SPACAL calorimeter is used to determine the energy 
of the scattered positron whereas the hadronic final state is measured by the
LAr calorimeter and the tracking systems.
The SPACAL calorimeter has an electromagnetic energy resolution
equal to $\sigma(E)/E=7.1\%/\sqrt{E [\mathrm{GeV}]} \oplus 1\%$
as measured in an electron beam~\cite{SPACALE1} and an angular acceptance of
$153^\circ < \theta < 177.5^\circ$.
The LAr calorimeter has a hadronic energy resolution equal to
$\sigma(E)/E \simeq 0.5/\sqrt E$ ($E$ in {\rm GeV}) as measured in a pion
beam~\cite{pionbeam} and  covers an angular range between
$4^\circ$ and $154^\circ$.
Charged track momenta are measured using the central jet chamber (CJC),
which consists of two concentric drift chambers covering an angular range
between $15^\circ$ and $165^\circ$, and by using the forward tracking system, 
which covers the angular range between $7^\circ$ and $25^\circ$.
A uniform field of $1.15 \ {\rm T}$ is
produced using a superconducting solenoid  which surrounds both the
LAr calorimeter and the central tracking system. Luminosity is measured by
detecting photons, from the Bethe--Heitler process
$e p \rightarrow e p \gamma$, in a crystal calorimeter situated at
$z = -103 \ {\rm m}$.

\subsection{Detection of Leading Protons}

\noindent
The final state proton data were collected during $1995$ 
with the H1 forward proton spectrometer~\cite{FPS-ICHEP96,blist}.
In order to measure momenta of protons with scattering angles 
$\theta \lsim 0.5$~{\rm mrad} with respect to the beam, the HERA machine 
magnets adjacent to the interaction region are employed as spectrometer 
elements. Protons which have lost more than $10$\% of their energy
in $e p$ interactions appear after about $80$~{\rm m} 
at a distance of several millimeters from the nominal orbit so that 
they can be registered in detectors close to the circulating beam.
The detector elements are mounted inside plunger vessels, so called 
Roman Pots, which are retracted during injection and are brought close to 
the beam after stable luminosity conditions are reached. The particles
detected by the FPS are assumed to be protons. This is in agreement
with Monte Carlo calculations which show that approximately $98$\% of 
the charged particles observed in the FPS are protons.

\begin{figure}[p]
\centering
\epsfig{file=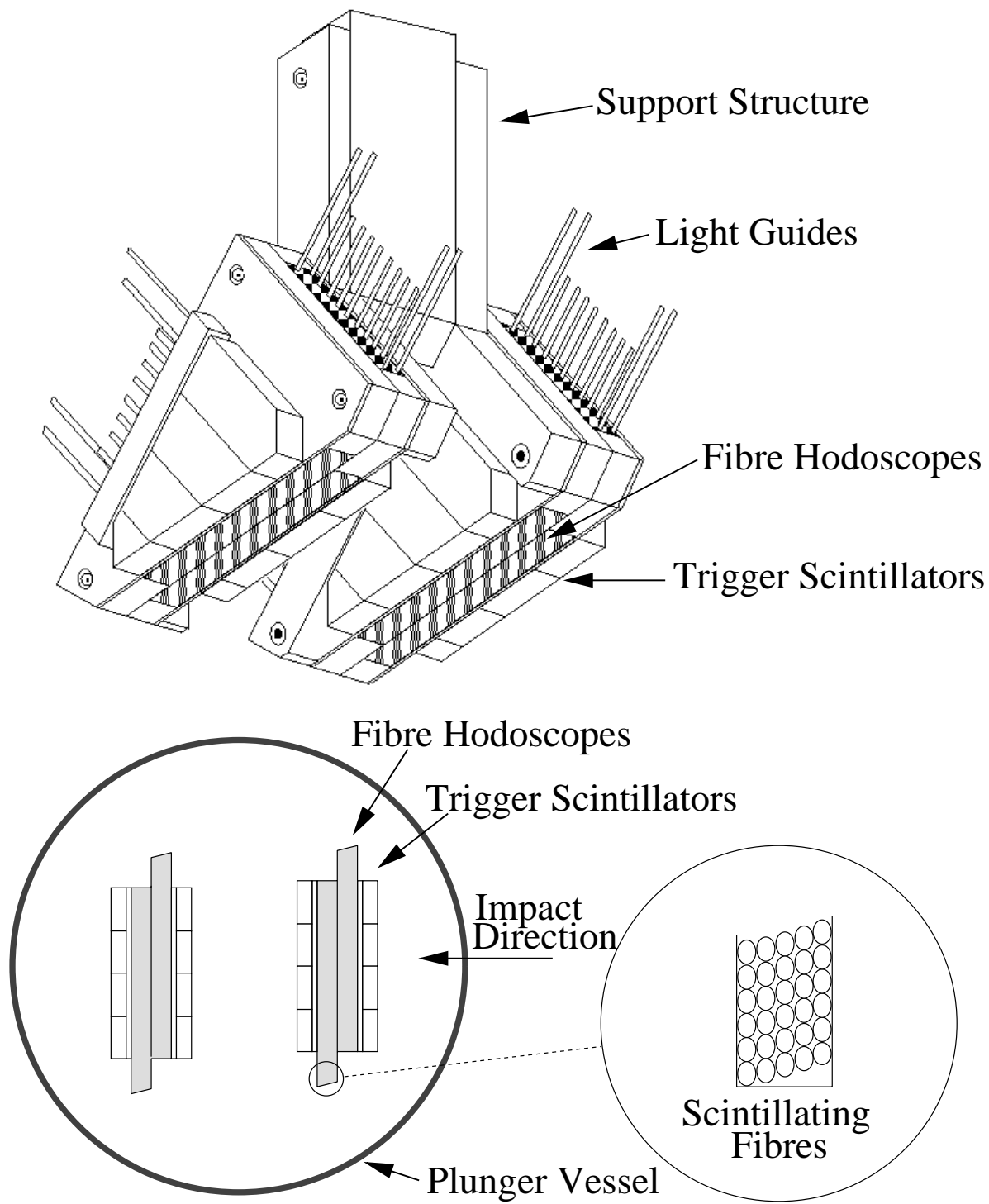,width=\textwidth}
\caption{Top: perspective view of the scintillating fibre hodoscopes and the 
trigger scintillators inside a Roman Pot. Bottom: horizontal cross section 
through one FPS station.}
 
\label{fpsview}
\end{figure}

During $1995$ H1 operated two FPS stations, located $81$ and $90$~{\rm m} 
away from the interaction point, which approach the beam from 
above. Each station is equipped with four planes of hodoscopes constructed of  
$1$~{\rm mm} scintillating fibres which are oriented $\pm 45^\circ$ 
with respect to the $y$ axis.
The scintillating fibre hodoscopes are 6~{\rm cm} wide and 2.5~{\rm cm} 
in height and are sandwiched between planes of trigger scintillators as 
sketched in Figure~\ref{fpsview}. Each scintillating fibre hodoscope has 
$240$ scintillating fibres arranged in five densely packed layers. Four of 
the scintillating fibres, belonging to a common layer but 
separated by $11$ fibres from each other, are attached to one cell of a 
$64$ pixel position sensitive photomultiplier (Hamamatsu H--4139--20).
Ambiguities in the hit combinations are resolved by using the segmented 
scintillators in front of and behind the hodoscopes.

The FPS detectors measure space points with a resolution of 
$\sigma\sub{x} = \sigma\sub{y} = 100\,\um$.
There is no magnetic field between the FPS detectors and the
space points measured at $81$ and $90$~{\rm m} are used to define 
a track at $85$~{\rm m}.
In both stations, ten out of the twenty layers have to show signals, 
in coincidence with trigger tiles, in order to reconstruct a 
track. The efficiency for a minimum ionizing particle to produce a hit in 
a layer is typically 60\% and the track reconstruction efficiency 
is approximately 50\%. With the help of the transfer matrices of 
the magnetic channel,
which are known with high accuracy between the interaction point and 
$85$~{\rm m}, trajectories are traced back to the interaction point and 
momenta are determined. Due to the fact that the HERA magnets lead to 
independent dispersion in both the horizontal and vertical planes, 
momenta can be measured 
twice by making use of the horizontal and vertical deflections. 
The two measurements have to agree within errors for tracks starting at 
the interaction point and this provides an efficient method for rejecting 
background tracks.

\subsection{Calibration of the Forward Proton Spectrometer}

\noindent
In order to reconstruct momenta, the coordinates and slopes 
of scattered proton trajectories are transformed into a reference 
system with the beam direction as the primary axis.
As the actual beam position is not known with the required accuracy of 
$0.5$~{\rm mm}, the actual beam orbit is determined for each fill using the
nominal orbit as the first approximation. First the offset and tilt of the 
actual beam orbit with respect to the nominal one is determined in the 
horizontal plane by a fit that makes use of the fact that certain 
combinations of impact points and slopes at 85~{\rm m} are ``forbidden'' for
particles coming from the nominal interaction vertex~\cite{FPS-ICHEP96}.
For the calibration of the momentum measurement in the vertical plane, the
difference between horizontal and vertical momentum measurements is used
as additional input. Only protons, for which the momentum error arising from 
the uncertainty of the calibration in the horizontal plane is small, are used 
for this purpose. This procedure, which has been verified by Monte Carlo 
simulations of the FPS, leads to a momentum resolution of typically 
$\sigma (E_p^\prime) = 6$~{\rm GeV} at $700$~{\rm GeV}. The angular resolution 
at the interaction point is $\sigma (\theta\sub{x}) = 5$~$\mu${\rm rad},  
while $\sigma (\theta\sub{y})$ varies between $5$ and $100$~$\mu${\rm rad}
depending upon energy and angle. The calibration of the FPS
is described in more detail in reference~\cite{blist}.

The FPS calibration was checked using high $Q^2$ DIS events with a 
forward rapidity gap~\cite{FPS-ICHEP96}. For events with these 
characteristics the hadronic final state is well
contained within the central H1 detectors and one can compare the observed 
missing longitudinal momentum with the one measured by the FPS
assuming that no particles escape in the forward region.
The mean difference between the proton energy measured in the FPS and
the energy expected from the calorimetric measurement is 
$(-1 \pm 9)$~{\rm GeV}. From this we conclude that the scale of the FPS
energy measurement is correct to within $10$~{\rm GeV}.

\subsection{Detection of Leading Neutrons}

\begin{figure}[p]
\centering
\epsfig{file=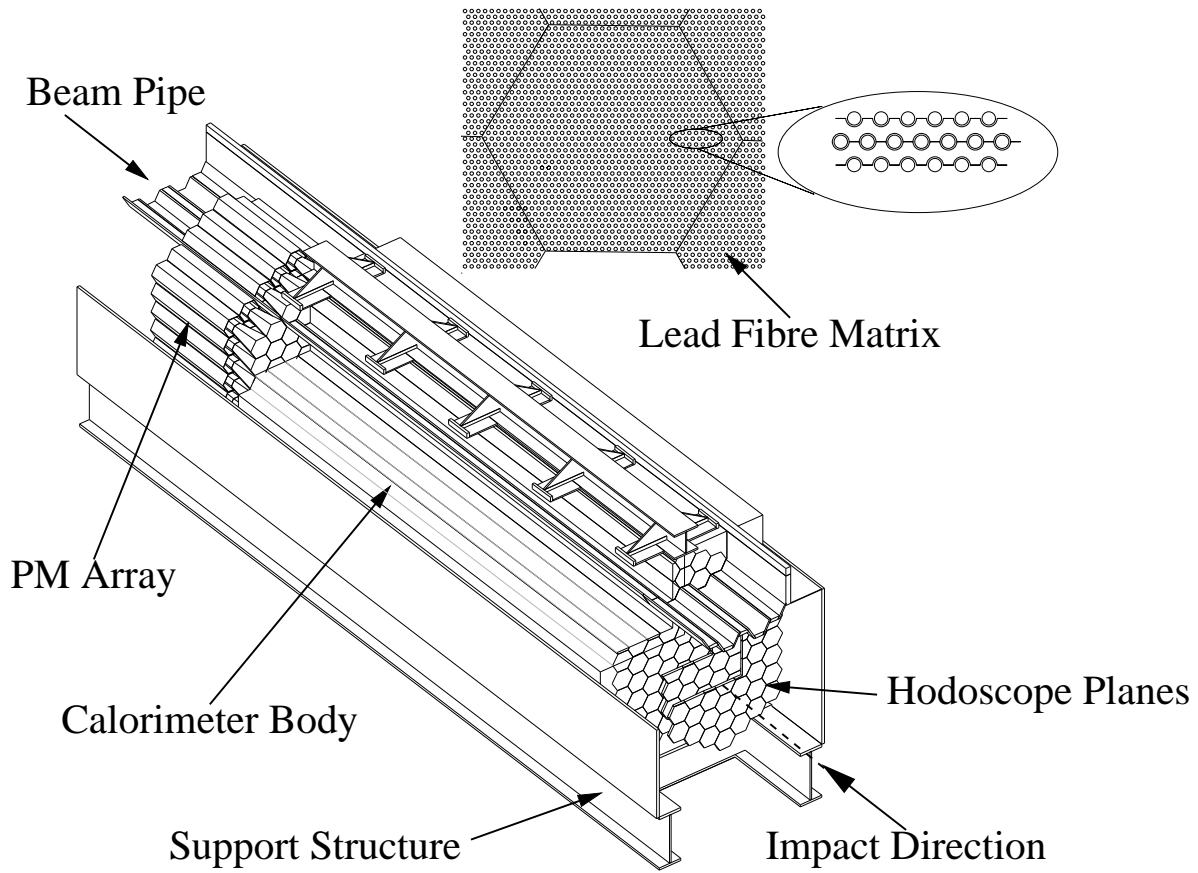,width=\textwidth}
\caption{Configuration of the H1 forward neutron calorimeter.
The calorimeter consists of interleaved layers of lead and
scintillating fibres. A hexagonal module, see inset, is defined by coupling
1141 scintillating fibres to a common photomultiplier located
at the rear of the detector.}
\label{FNCfig}
\end{figure}

\noindent
During 1996 the H1 experiment used a forward neutron calorimeter
constructed of lead and scintillating fibres. The calorimeter, which
was originally used by the WA89 experiment~\cite{cernlaa,WA89,BECK}
at CERN, weighs approximately 10 tons and is located 107~{\rm m} away from the 
nominal H1 interaction point. Final state neutrons with production angles
$\theta \lsim 0.5$~{\rm mrad} are within the acceptance of the FNC.
A schematic diagram of the FNC is shown in Figure~\ref{FNCfig}.

The forward neutron calorimeter consists of interleaved layers of 
2~{\rm m} long lead strips and scintillating fibres. The lead to fibre volume 
ratio is 4:1 and the nuclear interaction--length $\lambda_I$ is $21$~{\rm cm}.
The calorimeter is laterally segmented into hexagonal modules each of which 
is defined by coupling 1141 scintillating fibres to a common photomultiplier 
located at the rear of the detector. The height of a hexagonal module 
is 8.6~{\rm cm}. A gap between the top and bottom parts of the calorimeter is 
necessary in order to have space for the proton beam pipe which passes 
through the calorimeter. There are $67$ modules in the bottom part of the 
calorimeter and $8$ modules in the top part. 
 
The scintillating fibres are 1~{\rm mm} in diameter and are orientated 
approximately parallel to the direction of the incident neutron. 
The attenuation length of the scintillating fibres is $(1.7 \pm 0.2)$~m
which has been measured using muons from cosmic events. Detailed 
GEANT~\cite{GEANT} simulation studies have shown that this small attenuation 
length is responsible for the high--energy tail (see Figure~\ref{FNC-CAL})
that we observe in the neutron energy spectra. The high--energy tail is
due to fluctuations in the longitudinal shower profile which lead to energy 
depositions close to the back--end of the calorimeter. Because of the
small distance to the photomultipliers, the produced scintillation light
is attenuated less than normal leading to an over--estimation of the
incident particle's energy. The energy resolution of the calorimeter is 
$\sigma(E)/E \approx 20$\% for energies between $300$ and $800$~{\rm GeV}.

Two segmented planes of hodoscopes situated in front of the FNC are used
to veto charged particles. Each plane is constructed of 1~{\rm cm} thick 
hexagonal scintillator tiles which have the same lateral size as the 
calorimeter modules. The neutron detection efficiency of the FNC is 
$(93 \pm 5)$\%, the losses being due to coincidences in the veto 
hodoscopes which mostly originate from the back--scattering of charged 
particles produced during the neutron's hadronic shower. 
This efficiency was determined by measuring the rate of 
signals in the hodoscopes as a function of the radial distance away from 
the neutron impact position reconstructed in the calorimeter.
Extrapolating the rate of signals in the hodoscopes to the region close to 
the impact position, the probability due to back--scattering was estimated.

In this analysis we assume that all neutral clusters are produced by neutrons.
Using the LEPTO Monte Carlo program~\cite{Lepto65} and a GEANT~\cite{GEANT} 
simulation of the H1 beam line, we estimate that the background contribution 
due to other neutral particles (mostly photons and $K_L^0$) is $6$\% for 
events with $z > 0.2$. For $z >0.6$, the background contribution is $2$\%.

All detector components, including the calorimeter and the hodoscope planes,
are covered with lead sheets in order to shield them from synchrotron 
radiation. 

\subsection{Calibration of the Forward Neutron Calorimeter}

\noindent
The 75 modules of the FNC were initially calibrated at CERN using a 
10~{\rm GeV} incident electron beam. The FNC was positioned on a movable 
platform which allowed the response of each module to be measured separately.
Preliminary calibration constants for the entire FNC were then determined 
by a matrix inversion procedure~\cite{tpn}. After this initial calibration,
the FNC had an approximately uniform response independent of impact position.

After the calorimeter was installed in the H1 beam line, run--dependent 
calibration constants were determined every few weeks by comparing the 
high--energy spectrum of neutrons, observed in interactions between the 
proton beam and residual gas in the beam pipe, with the results of a 
$p p \rightarrow n X$ Monte Carlo simulation based upon pion 
exchange~\cite{tpn2}. 
Hadronic $p p \rightarrow n X$ data in the high $z$ and low $p_T$ range,
obtained at the ISR and by other experiments at CERN~\cite{data1,data2}, 
are well described by pion exchange and have 
been used to constrain the pion flux factor~\cite{Holtmann2,Holtmann}. 
Since the pion flux factor determines the high--energy spectrum of 
final state neutrons, by comparing with the Monte Carlo simulation we 
are effectively calibrating the FNC with respect to previous experimental 
results. In the $p p \rightarrow n X$ Monte Carlo program, the acceptance 
and the energy response of the FNC are simulated by tracking neutrons through
the GEANT~\cite{GEANT} simulation of the H1 beam line.

\begin{figure}[p]
\centering
\epsfig{file=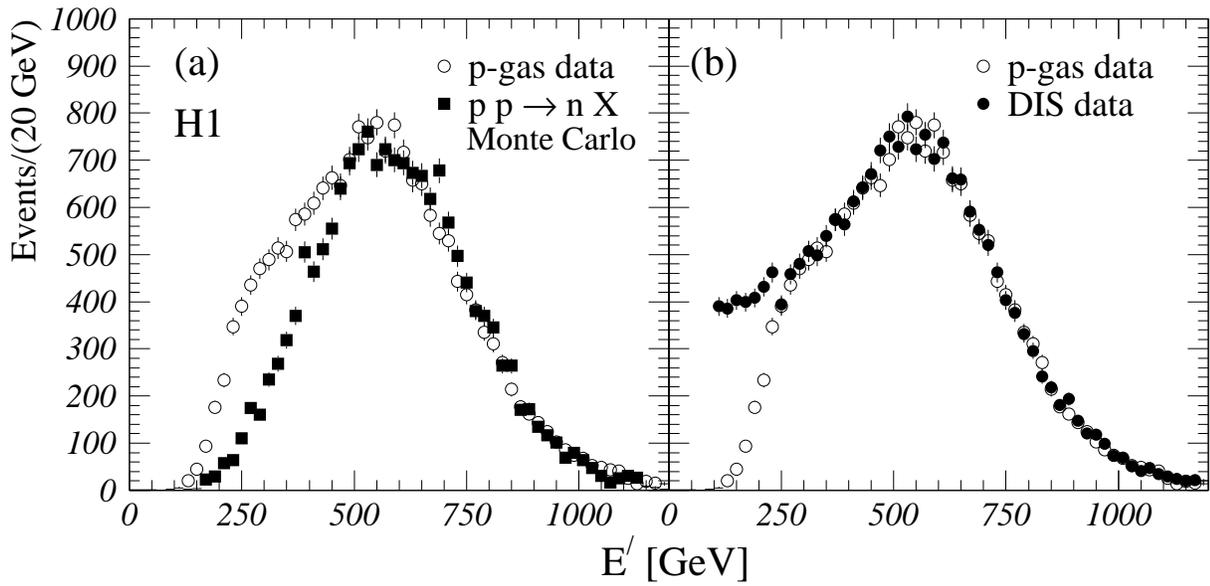,height=10cm}
\caption{(a) The observed neutron energy spectrum in proton beam--gas
interactions compared to the results of a $p p \rightarrow n X$
Monte Carlo simulation based upon pion exchange.
The Monte Carlo simulates the acceptance and response of the FNC. 
(b) The same proton beam--gas energy spectrum compared to the neutron 
energy distribution observed in DIS interactions. The proton beam--gas 
energy spectrum has not been corrected for the trigger efficiency which 
is less than 100\% below $300$~{\rm GeV}. All distributions are normalized 
to the number of events with $E^{\prime} \ge 500$~{\rm GeV}.}
\label{FNC-CAL}
\end{figure}

Figure~\ref{FNC-CAL}a shows the uncorrected neutron energy spectrum observed 
in proton beam--gas interactions compared to the results of the 
Monte Carlo simulation. The two distributions are normalized to the same 
number of entries above 500~{\rm GeV}. The peak position and the high--energy
tail observed in the data are in good agreement with the simulation.
Since the rate of neutron production with $z < 0.5$ is known to be 
underestimated by pion exchange~\cite{Holtmann2}, we do not use proton 
beam--gas data in this energy range for calibration purposes.
This comparison, between proton beam--gas interactions and the pion 
exchange Monte Carlo simulation, is the method we use to calibrate the FNC. 
We estimate a 5\% energy scale uncertainty for the FNC based upon our 
comparison between proton beam--gas interactions and the pion exchange 
Monte Carlo simulation. 

Figure~\ref{FNC-CAL}b shows the same beam--gas energy spectrum compared 
to the neutron energy spectrum observed in DIS interactions. Above 
$300$~{\rm GeV} the two distributions again agree very well in shape.
This agreement supports the hypothesis that the pion flux factor
is a universal property of the proton which is the same in both
DIS and hadronic interactions~\cite{Sullivan,Holtmann}.
Below $300$~{\rm GeV}, the sharp rise in the proton beam--gas energy
spectrum is due to the trigger threshold used to obtain the data.

The short--term gain variation of the FNC photomultipliers is measured by a 
LED monitoring system~\cite{tpn}. The light from seven 
light--emitting--diodes is coupled 
by optical fibres to the entrance windows of all the photomultipliers. 
The average response of the FNC photomultipliers to the LED light is 
used on a run--by--run basis to correct for small changes in the gain of the 
FNC photomultipliers. When there are stable beam conditions, 
the gain variation of the FNC photomultipliers is typically less than 
$0.1$\% during $30$ minutes.

The spatial resolution of the FNC was determined using charged particles
and three small scintillator counters situated in front of the calorimeter. 
The scintillators 
are $3~\times~3~\times~10$~{\rm mm}$^3$ and they are used in coincidence
with a hodoscope tile to define a trigger. The spatial resolution of the 
FNC was determined to be: 
\begin{equation}
       \sigma_{xy}(E) = {\bigg (} \frac{5.13 \pm 0.81}{\sqrt{E \, [{\rm GeV}]}}
       + (0.22 \pm 0.07) {\bigg )} \, {\rm cm},
\end{equation}
where the reconstructed impact position was determined using the 
centre--of--gravity of the hadronic shower and an empirical formula 
which corrects for the hexagonal shape of the calorimeter 
modules~\cite{Scheel,tpn}.

\section{Event Selection and Data Analysis}

\noindent
The final state proton and neutron data used to measure the 
semi--inclusive structure functions were collected during 
different years. The proton data were 
obtained during 1995 using a trigger which required a charged track through 
both detector stations of the FPS and a localized cluster 
in the backward (SPACAL) electromagnetic calorimeter. For part of the data 
a track candidate in the central jet chamber was also required by the trigger.
During 1996, a trigger which required an energy deposit in the SPACAL 
electromagnetic calorimeter and the absence of out--of--time background 
signals was used to obtain the DIS data containing a high--energy neutron. 

During the offline analysis selection, criteria were applied to the data
in order to reduce beam related backgrounds, events due to photoproduction
and events from reactions in which the incoming positron lost a significant 
amount of energy by radiation. The DIS selection criteria used
in the analysis are:

\begin{itemize}
\item  A positron with energy $ E^\prime_e \ge 12$~{\rm GeV}  in the 
       angular range $156^\circ \le \theta_e \le 177^\circ$ was required
       which ensures that the scattered positron is within
       the acceptance region of the SPACAL electromagnetic calorimeter.
\item  The DIS kinematic variables were required to be in the range
       $2 \le Q^2 \le 50$~{\rm GeV}$^2$, $0.02 \le y \le 0.6$ and
       $6\cdot 10^{-5} \le x \le 6\cdot 10^{-3}$.
       The kinematic variables are reconstructed using the $\Sigma$ method 
       \cite{Sigma} which combines the scattered positron energy and 
       angle measurements with the quantity $\Sigma$, which is the sum
       over all hadronic final state particles of the differences
       between energy and longitudinal momentum.
       The $\Sigma$ method has good resolution and 
       keeps radiative corrections small over the entire kinematic range
       considered here. 
\item  The quantity $\sum_i (E_i - p_{z,i})$, which is calculated
       using the energy $E_i$ and the longitudinal momentum $p_{z,i}$ of
       all final state particles including the scattered
       positron, is expected to be twice the electron beam energy. This
       quantity was required to be $\ge 40$~{\rm GeV}
       for the neutron analysis and $\ge 41.6$~{\rm GeV} for the proton
       analysis.
       These cuts suppress radiative events and photoproduction background.
\item  The reconstructed vertex position was required to be within
       $\pm 30$~{\rm cm} of the nominal vertex position in $z$.
\item  The proton analysis required at least one central track in 
       the CJC with $p_T \ge 450$~{\rm MeV} to originate from the 
       interaction vertex.
\end{itemize}

Cuts related to the final state baryons are:

\begin{itemize}
\item  For the proton data, one forward track with 
       $580 \le E_p^\prime \le 740$~{\rm GeV} and $p_T \le 200$~{\rm MeV} was 
       required to be detected by the FPS.
       Fiducial cuts on $\theta_x$ and $\theta_y$, which depended on the 
       proton energy, were applied to ensure that the track was observed in 
       a region of the phase space where the acceptance was well understood 
       and stable over the run period.
\item  For the neutron data, one neutral cluster with 
       $E_n^\prime \ge 100$~{\rm GeV} and $p_T \le 200$~{\rm MeV}
       was required to be reconstructed in the FNC. 
\end{itemize}

After these cuts the data samples were grouped into 12 $(x,Q^2)$ bins
in the range $6\cdot 10^{-5} \le x \le 6\cdot 10^{-3}$ and 
$2 \le Q^2 \le 50$~{\rm GeV}$^2$.
The proton data sample consists of 1661 events
and the neutron data consists of $10366$ events. The total luminosities 
of the proton and neutron data samples are $(1.44 \pm 0.03)$~pb$^{-1}$ and 
$(3.38 \pm 0.07)$~pb$^{-1}$ respectively.

The acceptances of the FPS and the FNC were determined by Monte Carlo
programs in which protons or neutrons from DIS reactions were tracked 
through a simulation of the HERA beam line. The finite aperture of the 
beam line magnets limits the acceptance of both the FPS and the FNC. 

\begin{figure}[p]
\centering
\epsfig{file=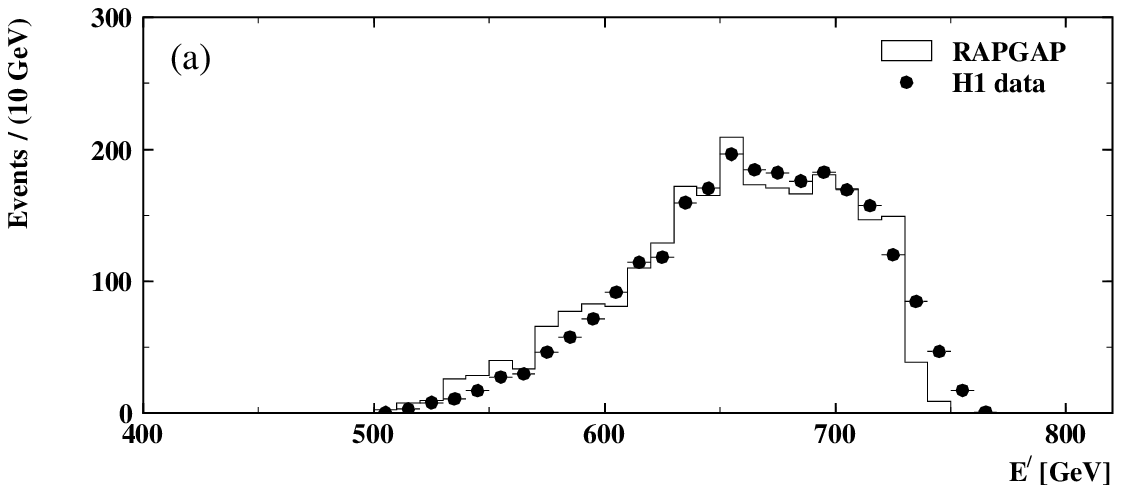,width=\textwidth}
\epsfig{file=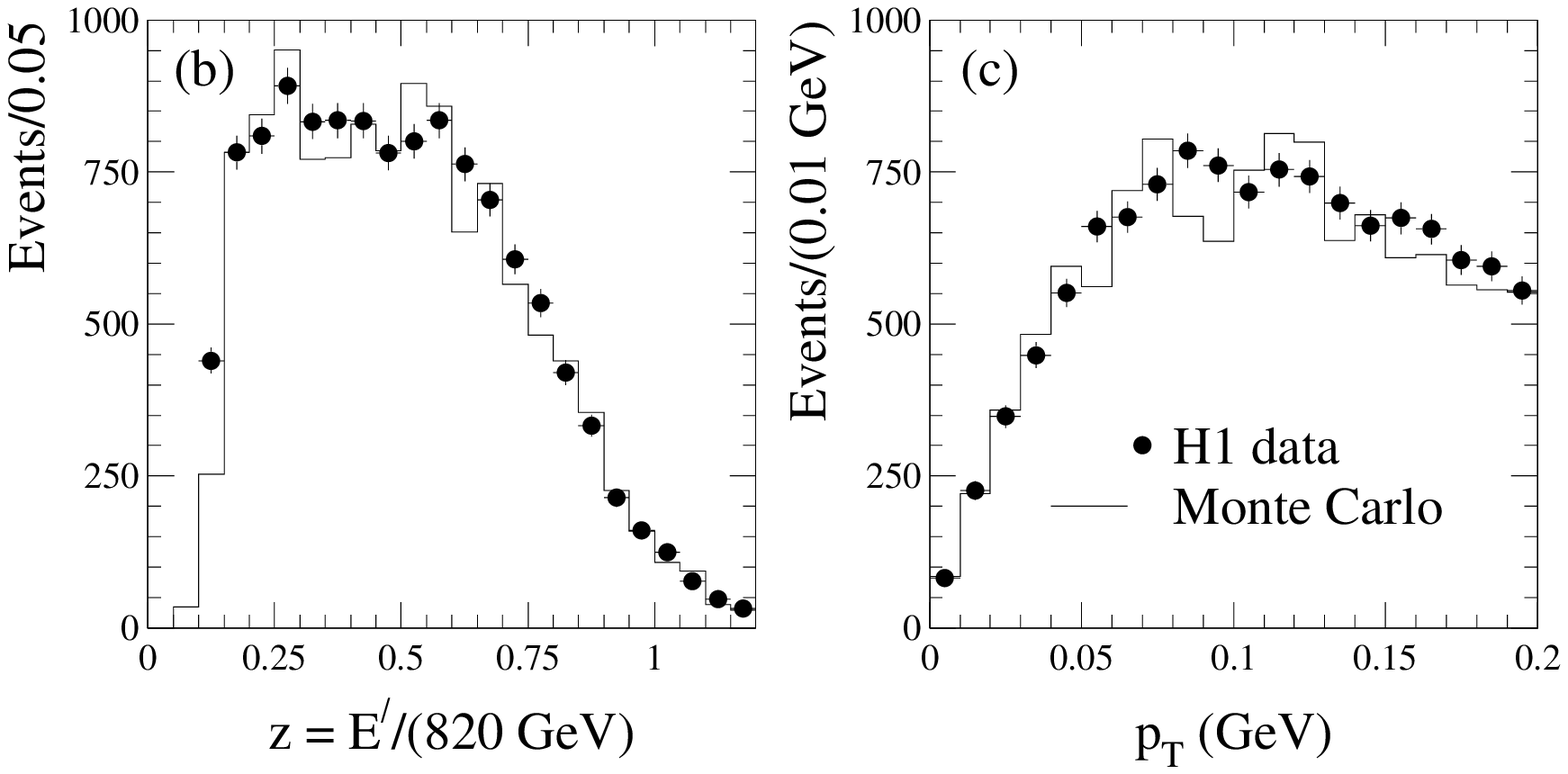,width=\textwidth}
\caption{(a) The observed proton energy spectrum compared to 
a simulation based upon the RAPGAP Monte Carlo generator.
In (b) and (c) the observed neutron $z$ and $p_T$ 
spectra, integrated over the entire kinematic range in $x$ and $Q^2$,
are compared to reweighted Monte Carlo data which result from the 
unfolding procedure used to correct the data for acceptance and
migration effects. The Monte Carlo distributions are 
normalized to the total number of events in the data.}
\label{FPS-FNC-MC}
\end{figure}

The FPS acceptance as a function of $z$ was calculated using the RAPGAP 
Monte Carlo program which simulates pion exchange~\cite{Rapgap}. This 
Monte Carlo simulation gives a good description of the shape of the 
uncorrected data as shown in Figure~\ref{FPS-FNC-MC}a for the
observed proton energy spectrum. The FPS acceptance
is approximately 80\% for protons with $0.76 \le z \le 0.90$ and 
$p_T \le 200$~{\rm MeV}. The LEPTO and ARIADNE~\cite{ARIADNE}
Monte Carlo programs were used to check that the FPS acceptance is independent 
of the assumed production model and to estimate the systematic uncertainties.

The corrected neutron energy spectrum was determined separately
in each $(x,Q^2)$ bin by using an unfolding procedure~\cite{Blobel}. The 
procedure uses Monte Carlo events to simultaneously correct the 
observed FNC energy spectrum for acceptance and migration effects.
The LEPTO and RAPGAP Monte Carlo models were used to demonstrate that the 
unfolded neutron energy spectrum does not depend upon the production model 
used to correct the data. For neutrons with $z \ge 0.4$ and 
$p_T \le 200$~{\rm MeV} the FNC acceptance is $\gsim 30$\%. 

Figures~\ref{FPS-FNC-MC}b and~\ref{FPS-FNC-MC}c show the observed neutron 
$z$ and $p_T$ spectra compared to the reweighted Monte Carlo 
simulation which results from the unfolding procedure. 
The data and Monte Carlo $z$ distributions, shown in Figure~\ref{FPS-FNC-MC}b,
are in agreement by construction since the unfolding procedure does a fit 
to the data by reweighting the Monte Carlo as a function of $z$.
The $p_T$ distributions, shown in Figure~\ref{FPS-FNC-MC}c, 
demonstrate that the reweighted Monte Carlo gives a good 
description of a variable not used in the fit.

The semi--inclusive structure functions ${\mathrm{F_2^{LP(3)}}}$ and 
${\mathrm{F_2^{LN(3)}}}$ have been corrected to the Born level.
Radiative corrections were calculated using the program 
HERACLES~\cite{HERACLES}. In all $(x,Q^2)$ bins they are smaller than 6\%. 
We have included a 2\% systematic error on our radiative corrections due to 
hadronic corrections and higher order processes which are not simulated by 
the HERACLES code.

There are three types of systematic errors: normalization
errors, errors which depend on the final state baryons and errors
which are different for each $(x,Q^2)$ bin:

\begin{itemize}
\item {Normalization Systematic Errors}
  \begin{itemize}
  \item For the proton analysis the normalization error is 
        5.6\%. The main contribution to this error is the 5.0\% uncertainty 
        in the proton reconstruction efficiency.
  \item The normalization error for the neutron analysis 
        is 5.7\%. The 5.4\% uncertainty in the neutron detection efficiency
        is the largest component of this error.
  \end{itemize}
  The systematic errors on the total integrated luminosities, which are 
  approximately $2$\%, are included in the normalization uncertainties.
\bigskip      
\item {Final State Baryon Systematic Errors}
  \begin{itemize}
  \item For the proton analysis, these errors are between 4.8\% and 19\%.
  Errors in the migration corrections for the proton energy intervals,
  which depend on the accuracy of the calibration procedure, were evaluated
  from Monte Carlo studies and range between 4.5\% and 19\% for the
  different bins. Additional errors due to
  the uncertainty in the acceptance of the the fiducial cuts were evaluated 
  by comparing the results obtained using the RAPGAP, LEPTO and ARIADNE 
  Monte Carlo models.
  \item 
  The FNC energy spectrum, after being corrected for acceptance and
  migration effects by using the unfolding procedure,
  has large systematic errors. We have varied the FNC energy
  scale by $\pm 5$\% and reweighted the Monte Carlo data as a function
  of $p_T$ to estimate these systematic errors. We have also used the
  shape of the impact point distribution and its maximum, which 
  defines the zero degree direction, to determine the FNC acceptance. 
  A systematic error is applied corresponding to the
  difference between this acceptance method and the acceptance
  determined using the unfolding procedure.
  These uncertainties lead to systematic errors which range between 16\%
  and 58\% for the corrected neutron energy spectrum.
  \end{itemize}
\bigskip  
\item {Errors which depend on $x$ and $Q^2$}
  \begin{itemize}  
  \item The systematic errors which differ in each $(x,Q^2)$ bin range 
        between $7.7$\% and $13$\% for the proton analysis. The uncertainties
        in the trigger efficiency and in the corrections for 
        migrations between different bins in $x$ and $Q^2$
        are the main contributions to these systematic errors.
        The uncertainty in the acceptance and in the migrations 
        as a function of 
        $x$ and $Q^2$ was determined by varying the energy scale of the 
        SPACAL calorimeter by $\pm 1.5$\%, by varying 
        the electron scattering angle by $\pm 1$~{\rm mrad} and by
        simulating, according to our knowledge of the hadronic energy scale 
        of the LAr calorimeter, a 4\% uncertainty in the measurement of 
        $\Sigma$.
  \item For the neutron analysis, the systematic errors which depend
        upon $x$ and $Q^2$ range between $3.1$\% and $7.2$\%, since the
        measurement of the scattered positron was further improved in 1996.
        The main source of these systematic errors is the
        uncertainty in the acceptance and in the migrations
        as a function of $x$ and $Q^2$ 
        which was determined by varying the energy scale of the SPACAL 
        calorimeter by between $\pm 1$\% and $\pm 3$\% (depending upon 
        the energy of the scattered positron),
        by varying the reconstructed angle of the scattered positron 
        by $\pm 0.5$~{\rm mrad} and  
        by varying the energy scale of the LAr calorimeter by $\pm 4$\%.
  \end{itemize}
\end{itemize}

These errors can be compared with the statistical ones. The statistical 
errors for the proton analysis lie between 9.6\% and 30\%
for 90\% of the data points. For the neutron analysis, the statistical
errors range between 4.7\% and 29\% for 90\% of the data points.

\section{\boldmath The Semi--Inclusive Structure Functions 
${\bf \mathrm{F_2^{LP(3)}}}$ and 
${\bf \mathrm{F_2^{LN(3)}}}$}
\label{results_section}

\begin{figure}[p]
\centering
\epsfig{file=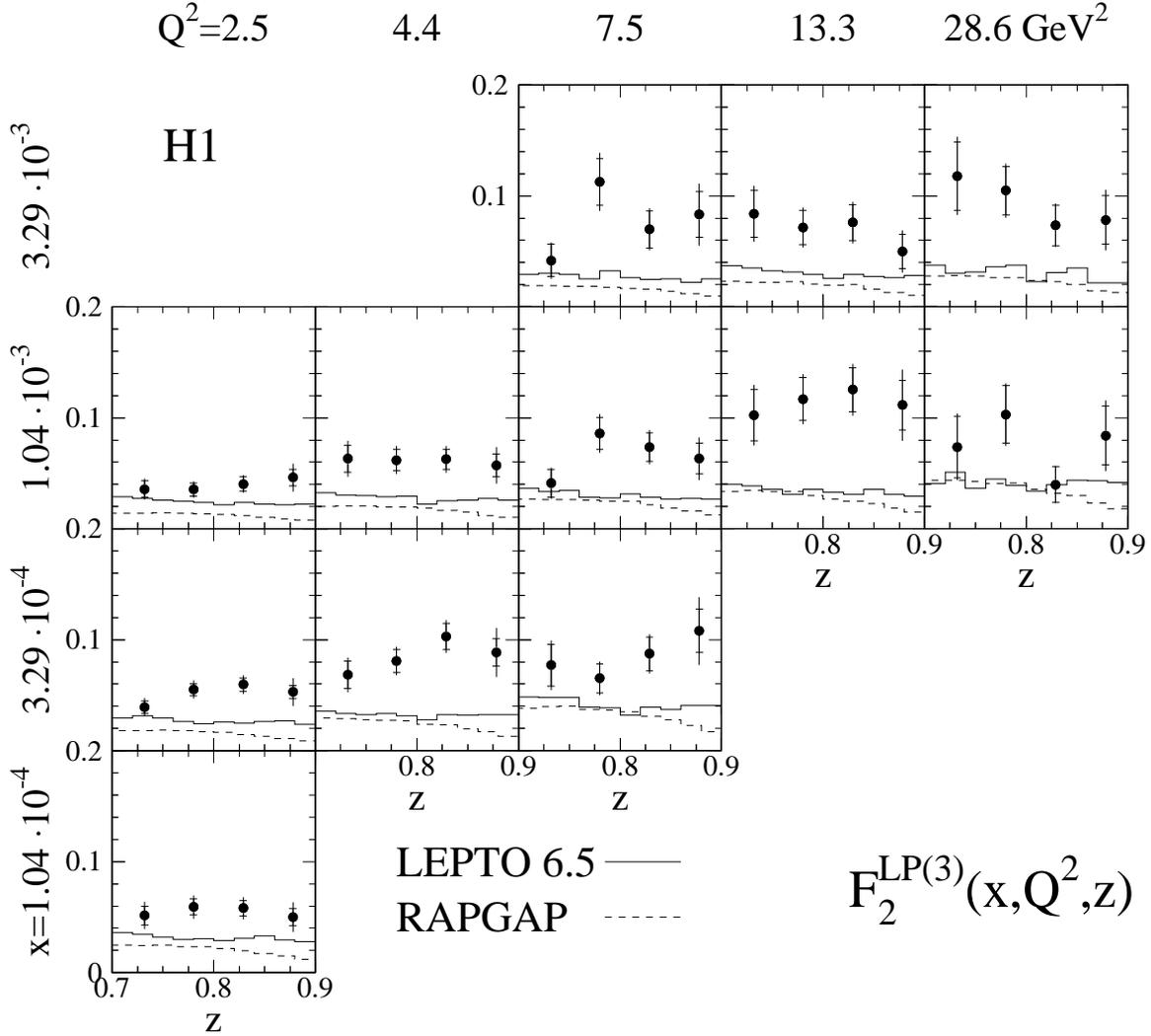,width=\textwidth}
\caption{Measurement of ${\mathrm{F_2^{LP(3)}}}$, for protons with
$p_T \le 200$~{\rm MeV}, compared to the predictions of the LEPTO and 
RAPGAP Monte Carlo models calculated using GRV leading order parton 
distributions for the proton and the pion respectively. The inner error 
bars show the statistical errors and the full error bars show the 
statistical and systematic errors added in quadrature. There is an 
additional 5.6\% overall normalization uncertainty for the data points 
which has not been included in the full error bars.}
\label{FPS-F2}
\end{figure}

\begin{figure}[p]
\centering
\epsfig{file=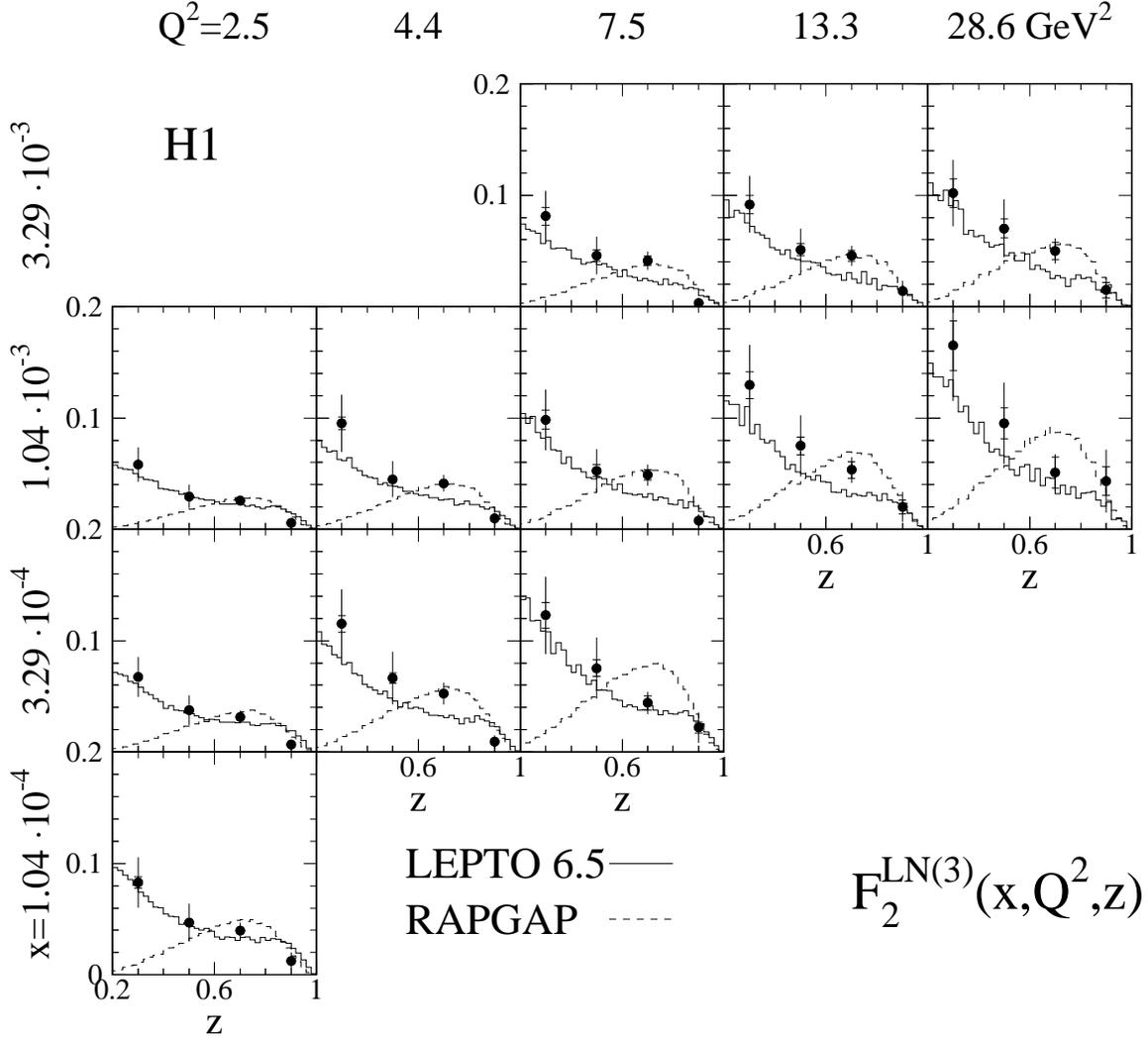,width=\textwidth}
\caption{The semi--inclusive structure function ${\mathrm{F_2^{LN(3)}}}$, for
neutrons with $p_T \le 200$~{\rm MeV}, compared to the predictions of the
LEPTO and RAPGAP Monte Carlo models calculated using GRV leading order parton 
distributions for the proton and the pion respectively. 
The inner error bars show the
statistical errors and the full error bars show the statistical 
and systematic errors added in quadrature. There is an additional 5.7\% 
overall normalization uncertainty for the data points which has not been
included in the full error bars.}
\label{FNC-F2}
\end{figure}

\noindent
Our measurement of the semi--inclusive structure functions 
${\mathrm{F_2^{LP(3)}}}$ and ${\mathrm{F_2^{LN(3)}}}$, for leading protons 
and neutrons with
$p_T \le 200$~{\rm MeV}, are shown in Figures~\ref{FPS-F2} and~\ref{FNC-F2}. 
The inner error bars show the statistical errors and the
full error bars show the statistical and systematic errors added in 
quadrature. 
Tables~\ref{f2lpdata} and~\ref{f2lndata} list the values 
of the semi--inclusive structure functions shown in the figures.
The data are compared to the predictions of the LEPTO and RAPGAP Monte 
Carlo models~\cite{Lepto65,Rapgap}. 

The LEPTO~6.5 Monte Carlo program simulates baryon production using 
soft colour interactions and the JETSET string fragmentation 
model~\cite{Lepto65,Jetset}. Soft colour interaction models have 
been proposed to explain large rapidity gap events and the production 
of final state baryons~\cite{Buch,Lepto-soft}. In these models, the 
colour structure of the partons interacting with the virtual photon is 
modified by non--perturbative soft gluon exchanges which
can lead to the production of colour neutral partonic subsystems 
separated in rapidity. After the fragmentation process, a high--energy baryon 
may be produced separated by a large rapidity gap from the remainder of the 
hadronic final state. 

The LEPTO Monte Carlo model describes the general shape and magnitude
of the neutron data over the entire $z$ range. It fails however to describe 
the rate of leading proton production and the rise in ${\mathrm{F_2^{LP(3)}}}$
as a function of $Q^2$. The leading order parton distributions for the 
proton by Gl\"uck, Reya and Vogt (GRV)~\cite{GRV,PDFLIB}, and the default
value of $0.5$ as the probability for soft colour interactions, were used
to calculate the LEPTO Monte Carlo predictions.

The RAPGAP Monte Carlo program simulates leading baryon production using 
pion exchange. In the Monte Carlo program, the cross sections for leading 
proton and neutron production are proportional to the product of the pion 
flux factor and the pion structure function. The pion flux factor 
determines the energy and $p_T$ spectra of the final state baryons and 
is identical for proton and neutron production except for a factor of two. 
For the RAPGAP Monte Carlo 
predictions shown in Figures~\ref{FPS-F2} and~\ref{FNC-F2} we have used 
the pion flux factor determined by Holtmann {\it et al.}~\cite{Holtmann}. 
The rate of leading baryon production depends also upon the values of the 
pion parton distributions and we have used the leading order 
parametrization by GRV~\cite{GRV,PDFLIB}. 

The RAPGAP Monte Carlo gives a reasonable description of the high--energy 
neutron data with $z \ge 0.7$ but it fails to reproduce the absolute rate 
of proton production. In the low--energy region 
where the final state neutron has $ < 70$\% of the 
incident proton's energy, the RAPGAP Monte Carlo program is not valid since 
additional physical processes, not simulated by the program, are expected 
to contribute significantly to the production of neutrons.

It is interesting to compare the magnitudes of 
${\mathrm{F_2^{LP(3)}}}$ and ${\mathrm{F_2^{LN(3)}}}$.
For $z \ge 0.7$, the semi--inclusive cross section for proton production
is larger than the cross section for neutron production in any specific
$(x,Q^2)$ bin. This result rules out pion exchange as the main production 
mechanism for leading protons since pion exchange models predict that 
the ratio of neutron and proton production should be equal to two.

\section{\boldmath Factorization and Scaling Violations of
${\bf \mathrm{F_2^{LP(3)}}}$ and ${\bf \mathrm{F_2^{LN(3)}}}$}
\label{factorization}

\noindent
Presuming that leading baryons emerge from reactions where the virtual
photon is absorbed by a colourless object inside the proton,
the structure function ${\mathrm{F_2^{LB(3)}}}$ should factorize
into a flux factor $f(z)$ which is only a function of $z$, and a structure
function ${\mathrm{F_2^{LB(2)}}}$ which depends upon $\beta$ and $Q^2$.
The quantity $\beta = x/(1-z)$ may be interpreted as the fraction of the
exchanged object's momentum carried by the quark or gluon interacting with
the virtual photon.

Alternatively one may assume models~\cite{Buch} which are not based on
the exchange of colourless objects so that $\beta$ can no longer be
interpreted as a momentum fraction. In such scenarios, one might expect
factorization in the variables $x$, $Q^2$ and $z$, if the deep--inelastic
scattering process off the proton is independent of the proton fragmentation.
The ``hypothesis of limiting fragmentation''~\cite{Chou}, which states that 
target fragmentation is independent of the incident projectile's energy, 
also implies that final state baryons emerge from a process which is 
insensitive to $x$ and $Q^2$.

To test both factorization hypotheses, fits were made to the proton and
neutron data separately assuming the following general forms for 
${\mathrm{F_2^{LB(3)}}}$~\cite{blist,tpn2}:
\begin{equation}
   {\mathrm{F_2^{LB(3)}}}(\beta,\, Q^2,\, z) = f(z) \cdot
   {\mathrm{F_2^{LB(2)}}}(\beta,\,Q^2)
\label{fit1}
\end{equation}
\begin{equation}
   {\mathrm{F_2^{LB(3)}}} (x,\, Q^2, \, z) = f(z) \cdot
   {\mathrm{F_2^{LB(2)}}}(x,\,Q^2),
\label{fit2}
\end{equation}
where the discrete--function $f(z)$ is expressed by three free--parameters
and for ${\mathrm{F_2^{LB(2)}}}(\beta,\,Q^2)$ in equation~\ref{fit1}
a functional form, based on the leading terms of a phenomenological
parametrization of the proton structure function~\cite{H1F2}, was chosen.
For ${\mathrm{F_2^{LB(2)}}}(x,\,Q^2)$ in equation~\ref{fit2},
$\beta$ was replaced by $x$.

The data are consistent with both factorization hypotheses and the
fit results yield similar $\chi^2/{\rm ndf}$.
A possible explanation for this result is that ${\mathrm{F_2^{LB(2)}}}$
is proportional to the proton structure function which 
for $x < 0.1$ is of the form 
${\mathrm{F_2}} \sim x^{-\lambda(Q^2)}$~\cite{H1F2}.
Since $\beta$ and $x$ are highly correlated and have similar magnitude
due to the restricted range of $z$, this also implies that
${\mathrm{F_2^{LB(2)}}} \sim \beta^{-\lambda(Q^2)}$.
The data have therefore relatively limited sensitivity 
to a difference of factorization in these two variables.

\begin{figure}
\centering
\epsfig{file=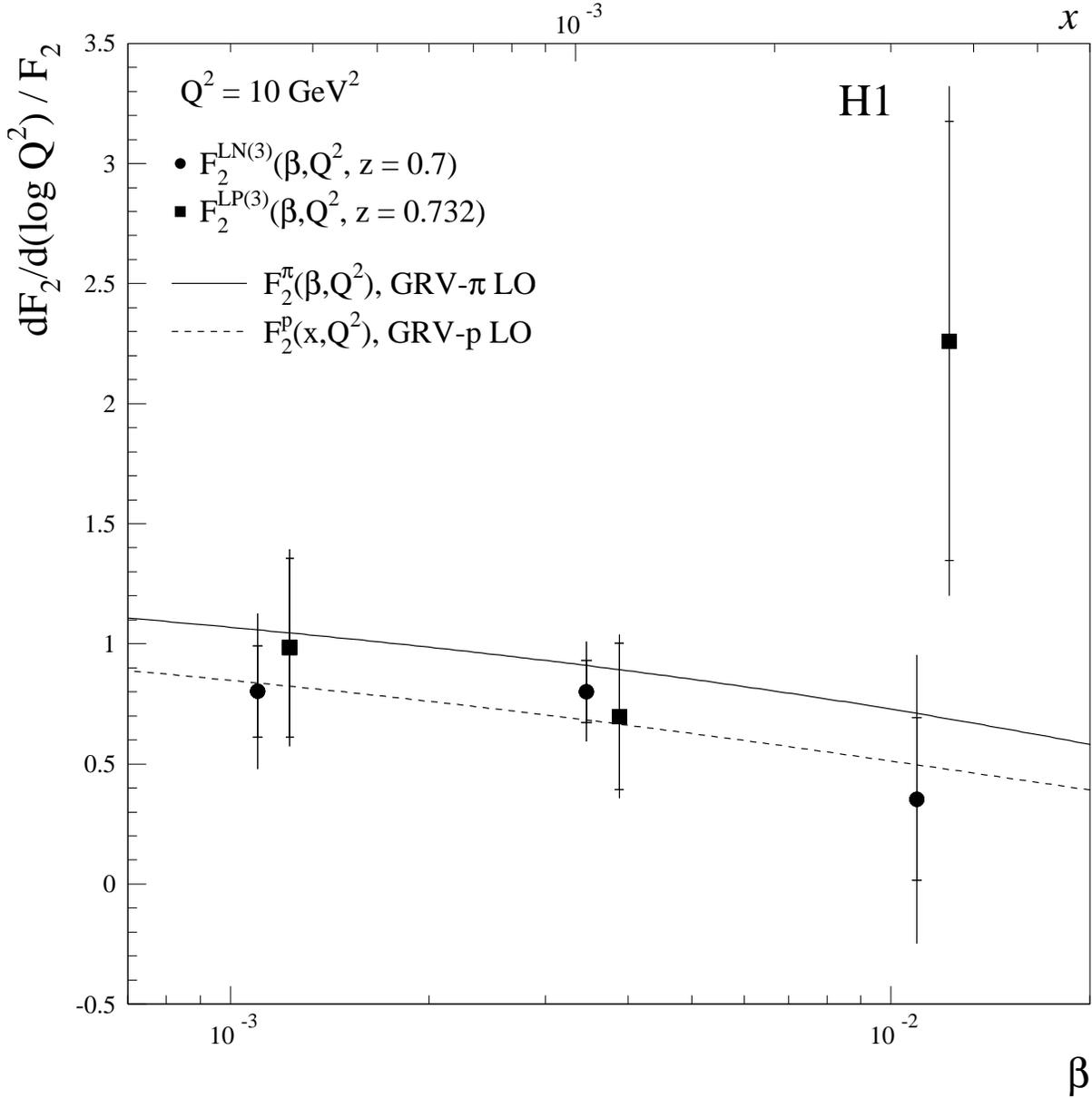,width=\textwidth}
\caption{Normalized scaling violations observed in the proton and
neutron data, computed at $Q^2 = 10$~{\rm GeV}$^2$, compared to
the expectations derived from GRV parametrizations of the inclusive
structure functions for the pion and the proton. The scaling
violations of the pion structure function have been evaluated as a function
of $\beta$ (lower scale), whereas for the proton structure function they
have been evaluated as a function of $x$ (upper scale).}
\label{scalebreak}
\end{figure}

In order to quantify the scaling violations observed in the data,
${\mathrm{F_2^{LP(3)}}}$ and ${\mathrm{F_2^{LN(3)}}}$ have been fitted 
separately for each fixed value of $\beta$ to the form:
\begin{equation}
  {\mathrm{F_2^{LB(2)}}}(\beta,Q^2) =a(\beta) + b(\beta) \cdot 
  \log{Q^2},
\end{equation}
with $Q^2$ in {\rm GeV}$^2$.
The values of $b(\beta)/{\mathrm{F_2^{LB(2)}}}$, which are a measure of 
the scaling 
violations, are plotted in Figure~\ref{scalebreak}. Only the fit results
which arise from the proton data with $z=0.732$ and the neutron data with 
$z=0.7$ are shown since the results from the other $z$ values are similar.
The measurements of ${\mathrm{F_2^{LP(3)}}}$ and ${\mathrm{F_2^{LN(3)}}}$ 
in the lowest $x$ bin have not been used since there is only a single 
$Q^2$ value. The data are compared to the scaling violations 
$\mathrm{d} {\mathrm{F_2}}/\mathrm{d} (\log Q^2)/{\mathrm{F_2}}$
predicted and observed in the inclusive structure functions of the pion and 
proton respectively.
The pion and proton structure functions have been calculated using the
GRV leading order parametrizations~\cite{GRV,PDFLIB}.
The scaling violations observed in the semi--inclusive structure functions 
${\mathrm{F_2^{LP(3)}}}$ and ${\mathrm{F_2^{LN(3)}}}$ are similar in 
size and shape and are close to those seen in the GRV parametrizations
of the pion and proton structure functions.

\section{Comparison to a Regge Model of Baryon Production}

\noindent
Assuming a simple Regge expansion and the dominance of a single Regge
exchange, the differential cross section for leading baryon production as
a function of $z$ at fixed $t$ should be proportional to $(1-z)^{-n}$.
Here $n=2\alpha(t)-1$, and $\alpha(t)$ specifies the Regge trajectory of
the dominant exchange. For the leading neutron data 
with $0.7 \le z \le 0.9$ shown in Figure~\ref{FNC-F2},
the falling $z$ spectra suggest a value of $n$, averaged over 
the $t$--dependence of the baryon production cross section,
which is approximately equal to $-1$. This implies that the average value 
of $\alpha(t)$ is consistent with zero which is naively the expectation of 
pion exchange. In contrast the leading proton data, shown in 
Figure~\ref{FPS-F2}, do not depend strongly 
on $z$ so that the average value of  $\alpha(t)$ is larger than 
the value suggested by the
neutron data. This is consistent with the dominance of a 
trajectory with the intercept $\alpha(0)\simeq 0.5$ which was
found to be the sub--leading contribution in the diffractive region at larger
$z$~\cite{H1F2D32}. 

\begin{figure}[p]
\centering
\epsfig{file=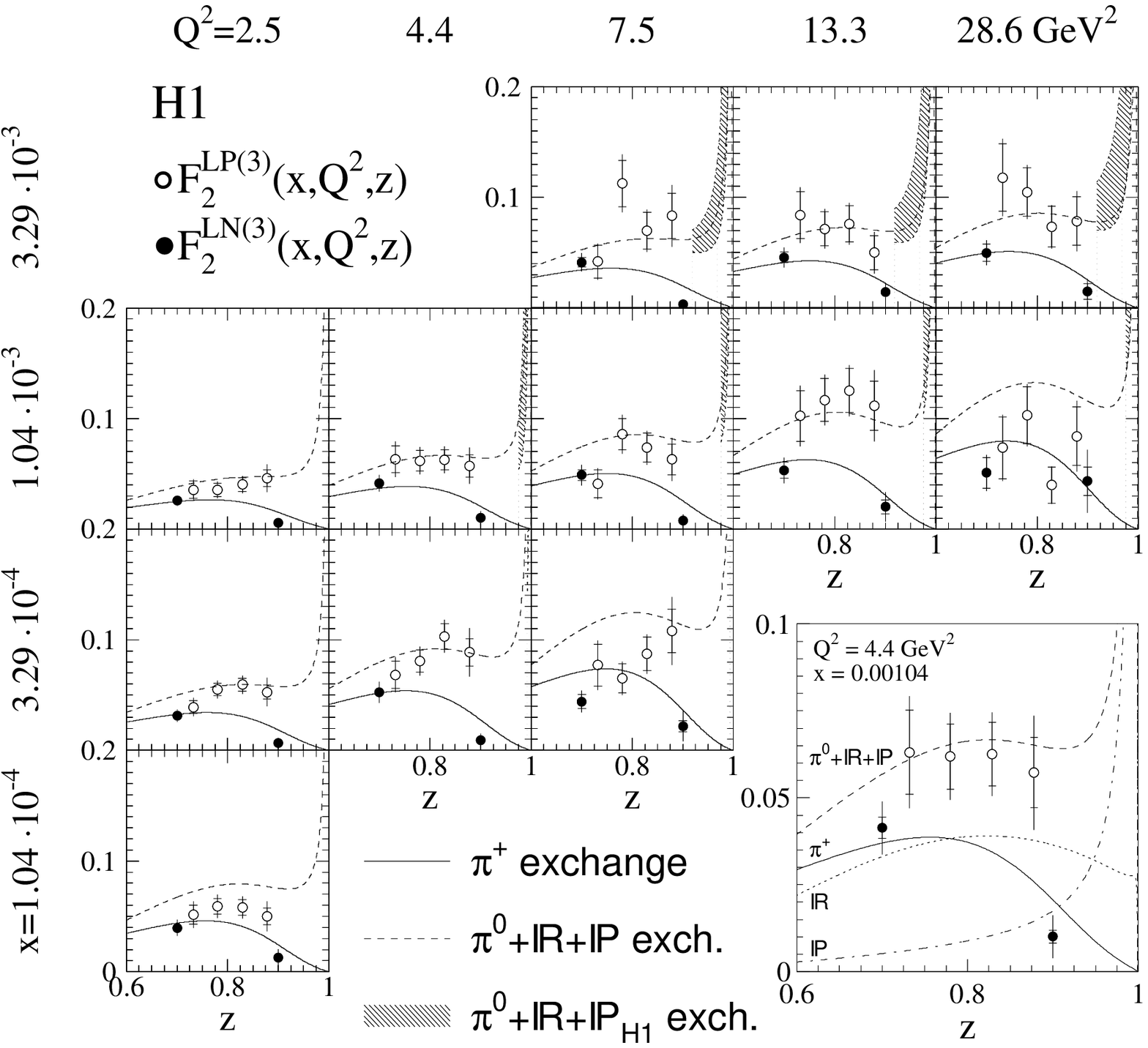,width=\textwidth}
\caption{The measured values of ${\mathrm{F_2^{LN(3)}}}$ and 
${\mathrm{F_2^{LP(3)}}}$ with $z \ge 0.7$ 
compared to a Regge model of baryon production.
The different contributions 
are labeled for the figure in the inset. The neutron data are 
described by $\pi^+$ exchange whereas the proton data are compared to 
the sum of $\pi^0$, pomeron and secondary reggeon $(f_2)$ exchanges.
The $\pi^0$ contribution, which is not shown, is exactly half the $\pi^+$ 
contribution. The shaded--band is explained in the text.}
\label{FNC-FPS}
\end{figure}

Figure~\ref{FNC-FPS} shows a comparison between the 
leading baryon structure functions with $0.7 \le z \le 0.9$ and a Regge 
model of baryon production. In the model, the contribution of a 
specific exchange $i$ is determined by the product of its particle 
flux $f_{i/p}(z,t)$ and its structure function $F_2^i$ evaluated at 
$(\beta,Q^2)$. For leading baryon production with $p_T \le 200$~{\rm MeV} we 
therefore have:
\begin{equation}
    {\mathrm{F_2^{LB(3)}}}(\beta,Q^2,z)= \sum_i {\Bigg (} 
    \int_{t_0}^{t_{\mathrm{min}}} f_{i/p}(z,t) \, {\mathrm d}t {\Bigg )} 
    \cdot {\mathrm{F_2^i}}(\beta,Q^2),
\label{mastereq}
\end{equation}
where $i$ denotes the pion, the pomeron and secondary reggeons
(for example $\rho$, $\omega$, $a_2$ and $f_2$).
The integration limits $t_0$ and $t_{\mathrm{min}}$ are given by 
equation~\ref{tlimits}.

In the Regge model, we assume that the neutral pion, the pomeron and 
the $f_2$ all contribute to leading proton production. We neglect the
contributions due to the other secondary reggeons because 
there is no sensitivity to them in the data, and because
they have been estimated to be much smaller than the contribution 
due to $f_2$ exchange~\cite{Golec,Kazarinov}. A comparison
of total hadronic cross section measurements has resulted in the estimate that 
the flux of reggeons which have isospin 
equal to one ($\rho$ and $a_2$) is only 
$\approx 3$\% of the flux of reggeons with isospin equal to zero 
($\omega$ and $f_2$)~\cite{Golec}. Regge phenomenology also predicts $f_2$ 
dominance, among isoscalar trajectories in the present case, 
in contrast to exchange degeneracy for elastic scattering 
processes~\cite{Kazarinov}.

For leading neutrons, we assume that they are produced by charged pion 
exchange only. In the limited $p_T$ range of the data, leading neutron 
production due to $\rho$ and $a_2$ exchanges has been estimated to be more 
than an order of magnitude smaller than the contribution due to pion 
exchange~\cite{Povh}. Pomeron exchange also does not give a significant 
contribution since neutron production due to diffractive dissociation is
believed to be $\approx 6$\% of the pion exchange contribution~\cite{Povh}. 
The present data sample has been used to estimate a 
$2$\% diffractive dissociation contribution to leading neutron production 
by determining the fraction of events with a large rapidity gap extending 
into the LAr calorimeter. We have neglected additional backgrounds such as 
neutron production due to resonance decays.
 
The pion, pomeron and reggeon flux factors have been determined using
hadron--hadron data. The pion flux factor $f_{\pi/p}$ which we have used for 
neutron production is the same as the one used in reference~\cite{Povh}:
\begin{equation}
   f_{\pi/p}(z,t) = C \frac {3 g_{\pi N p}^2}{16 \pi^2} \, 
   (1-z)^{1-2 \alpha_\pi^\prime t} \,
   \frac{|t|}{(m_\pi^2-t)^2}
   \, {\mathrm{exp}} {\Bigg (} 2 R_\pi^2 (t-m_\pi^2) {\Bigg )}, 
\label{pion-flux}
\end{equation}
where $g_{\pi N p}^2/(4\pi) = 13.6 \pm 0.3$~\cite{Timmermanns},
$\alpha_\pi^\prime =1$~{\rm GeV}$^{-2}$, $R_\pi^2 = 0.3$~{\rm GeV}$^{-2}$ 
and the square of the Clebsch--Gordon 
coefficient is $C = 2/3$. For proton production via $\pi^0$ exchange we 
use the same flux factor with $C = 1/3$. The pomeron and reggeon 
flux factors are parameterized as \cite{Szczurek,Golec}:
\begin{equation}
  f_{\pom/p} (z,t) = \frac {54.4 \, {\mathrm{GeV}}^{-2}}{8 \pi^2} \,
   (1-z)^{1-2 \alpha_\pom(t)} \,
   {\mathrm{exp}} {\Bigg (} 2 R_\pom^2 t {\Bigg )}
\end{equation}
\begin{equation}
   f_{\reg/p} (z,t) = \frac {390 \, {\mathrm{GeV}}^{-2}}{8 \pi^2} \, 
   (1-z)^{1-2 \alpha_\reg(t)} \,
   {\mathrm{exp}} {\Bigg (} 2 R_\reg^2 t {\Bigg )}, 
\end{equation}
where $\alpha_\pom(t)=(1.08 + 0.25 \, {\mathrm{GeV}}^{-2} \, t)$ 
and $\alpha_\reg(t) = (0.5 + 0.9 \, {\mathrm{GeV}}^{-2} \, t)$.
The slopes are $R_\pom^2 = 1.9 \, {\mathrm{GeV}}^{-2}$  and
$R_\reg^2 = 2 \, {\mathrm{GeV}}^{-2}$ respectively.
The modulus squared of the reggeon signature factor~\footnote
{In reference~\cite{Szczurek}, equation 7 is missing 
the reggeon signature factor which is given in equation 5 of 
reference~\cite{Golec}. 
The two publications also use different values of $R_\reg^2$.
In reference~\cite{Golec}, $R_\reg^2 = 1.2$~{\rm GeV}$^{-2}$ which leads 
to a 12\% difference in the values of the $p_T$--integrated reggeon flux 
factor at $z=0.8$.}, which is approximately equal to two, has been 
absorbed into the reggeon coupling and we have not included reggeon--pomeron 
interference terms in the model.

The evaluation of the pion flux factor is not without some
theoretical uncertainty. It has been pointed out that absorptive 
corrections, generated by double reggeon pion--pomeron exchanges,
might play an important role in hadronic reactions in contrast 
to DIS. Since the pion flux factor which we have used was determined using 
$p p \rightarrow n X$ data, it might underestimate the flux of 
pions in the proton for DIS reactions by up to $\approx 
30\%$~\cite{Nik97,Pirner}. 

The structure functions for the exchanged particles are basically unknown 
in the low $\beta$ region and one has to rely on theoretical models. For 
the pion structure function ${\mathrm{F_2^\pi}}$ we took the leading order 
parametrization by Gl\"uck, Reya and Vogt~\cite{GRV,PDFLIB}. For the reggeon
and pomeron structure functions we assume ${\mathrm{F_2^\reg}}
={\mathrm{F_2^\pi}}$ and 
${\mathrm{F_2^\pom}}=(0.026/0.12) 
{\mathrm{F_2^\reg}}$ following the arguments given in
reference~\cite{Szczurek}. Measurements of the diffractive structure 
function ${\mathrm{F_2^{D(3)}}}$~\cite{H1F2D3,ZEUSF2D3,H1F2D32,ZEUSF2D32}
only probe the pomeron at high $\beta$ $(\beta > 0.04)$
and it is not possible to use these data to fix the pomeron structure 
function ${\mathrm{F_2^\pom}}$ at the low $\beta$ values of the
semi--inclusive data $(\beta < 3\cdot 10^{-3})$. In the small region
of overlap however, the QCD fits to ${\mathrm{F_2^{D(3)}}}$~\cite{H1F2D3}
are consistent with the pomeron model used in this paper as will be 
discussed below.

The model gives an acceptable description of the neutron and proton data with
$0.7 \le z \le 0.9$, in view of the fact that all particle fluxes and
structure functions were taken from the literature and that no adjustment
was made. The rate of leading neutron production can be described 
entirely by $\pi^+$ exchange. However, proton production requires 
contributions from both $f_2$ and $\pi^0$ exchange which are roughly in 
the ratio $2:1$ from the model.

The shaded--band in Figure~\ref{FNC-FPS} shows the prediction for
${\mathrm{F_2^{LP(3)}}}$ in which we have replaced the pomeron
component in our Regge model with the pomeron component determined
using the QCD fit to ${\mathrm{F_2^{D(3)}}}$~\cite{H1F2D3}. 
In the QCD fit,
the pomeron structure function is parameterized at a low scale 
and evolved to larger $Q^2$ using the leading order DGLAP~\cite{DGLAP} 
equations. 
The hard--gluon leading order result which we have used for
${\mathrm{F_2^\pom}}$ (fit 3 in reference~\cite{H1F2D32})
is only shown in the region in which it is valid
($3 \le Q^2 \le 75$~{\rm GeV}$^2$ and $0.04 \le \beta \le 1.0$) and it
has been interpolated from 
$|t| \le 1$~{\rm GeV}$^2$ to $p_T \le 200$~{\rm MeV} in order to allow 
comparison with the leading proton data.
The width of the band reflects the uncertainty in the interpolation
to the different kinematic region.
The pomeron flux factor used in reference~\cite{H1F2D3}
has been evaluated
using $\alpha_\pom(t)=\alpha_\pom(0) + \alpha_\pom^\prime \, t$, where
$\alpha_\pom(0) =1.203 \pm 0.020 \, {\mathrm{(stat.)}} \pm 0.013 \,
{\mathrm{(syst.)}}$ and
$\alpha_\pom^\prime = (0.26 \pm 0.26)$~{\rm GeV}$^{-2}$~\cite{H1F2D3}.
The ZEUS measurement of the slope parameter
$b = (7.2 \pm 1.1 \, {\mathrm{(stat.)}}^{+0.7}_{-0.9} \,
{\mathrm{(syst.)}})$~{\rm GeV}$^{-2}$~\cite{ZEUSSLOPE},
where $b = 2 R_\pom^2 - 2 \alpha_\pom^\prime \ln{(1-z)}$,
has also been used.
This comparison demonstrates that the
H1 measurements of ${\mathrm{F_2^{D(3)}}}$ and ${\mathrm{F_2^{LP(3)}}}$ 
can both be described by Regge phenomenology.

\begin{figure}[p]
\centering
\epsfig{file=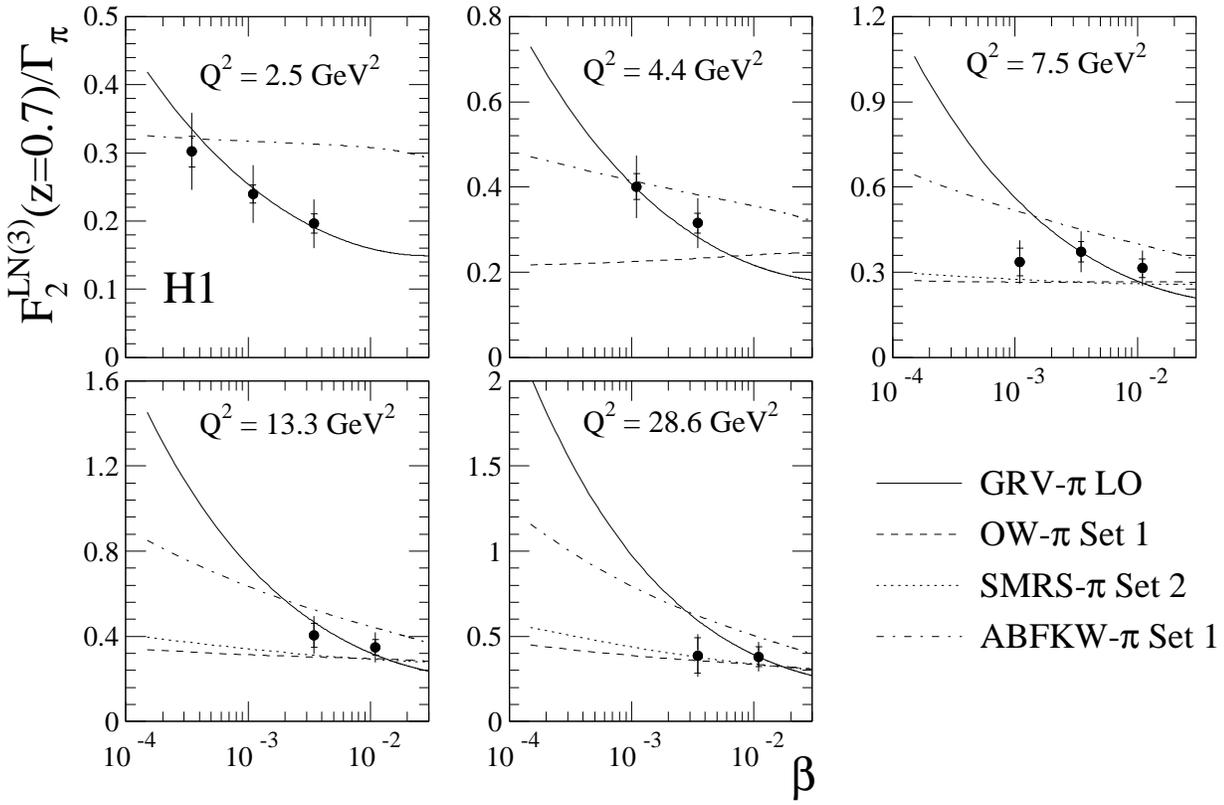,width=\textwidth}
\caption{${\mathrm{F_2^{LN(3)}}}/\Gamma_{\pi}$ at $z=0.7$ plotted 
as a function of 
$\beta$ for fixed values of $Q^2$. The quantity $\Gamma_{\pi}$
is the $p_T$ integrated pion flux factor.
Within the framework of the Regge model, ${\mathrm{F_2^{LN(3)}}}/\Gamma_{\pi}$ 
can be interpreted as being equal to the pion structure function 
${\mathrm{F_2^\pi}}$.
The data are compared to different parametrizations of 
${\mathrm{F_2^\pi}}$ which
are only shown in the $Q^2$ regions in which they are valid.}
\label{FNC-F2PI}
\end{figure}

We use the measurement of ${\mathrm{F_2^{LN(3)}}}$ at 
$z=0.7$ and the integral of the 
pion flux factor to estimate the pion structure function at low
Bjorken--$x$. Assuming
that our Regge model of leading neutron production is valid, the quantity 
${\mathrm{F_2^{LN(3)}}}/\Gamma_{\pi}$ can be 
interpreted as being equal to the structure function of the pion where:
\begin{equation}
   \Gamma_{\pi}(z=0.7) = \int_{t_0}^{t_{\mathrm{min}} }
      f_{\pi/p}(z=0.7,t) \, {\rm d}t = 0.131.
\end{equation}
Figure~\ref{FNC-F2PI} shows ${\mathrm{F_2^{LN(3)}}}/\Gamma_{\pi}$ as a 
function of $\beta$ for fixed values of $Q^2$. 
The data are compared to predictions of several parametrizations of the 
pion structure function~\cite{GRV,PDFLIB,OWENS,AURENCHE,SMRS}.
The latter are only shown in the $Q^2$ regions in which they are valid.
The data are in good agreement with the expectations of the 
GRV leading order parametrization of the pion structure function.

The quark and gluon distributions of the pion have previously been
constrained in the $x \gsim 0.1$ region using Drell--Yan data and direct
photon production data obtained by $\pi p$ scattering 
experiments (see for example \cite{Conway,NA3,NA10,WA70,NA24}).
Our determination using ${\mathrm{F_2^{LN(3)}}}$ is the first result
which constrains the pion structure function at values of $x$ which are 
more than an order of magnitude smaller. Background contributions and 
possible absorptive corrections~\cite{Povh,Nik97,Pirner}, which
have not been taken into account, are expected to only affect the absolute 
normalization of our result since all of the data are at $z = 0.7$.

\section{Summary and Conclusions}

\noindent
The semi--inclusive cross sections $e p \rightarrow e p X$ and
$e p \rightarrow e n X$ have been measured in the 
kinematic region 
$2 \le Q^2 \le 50$~{\rm GeV}$^2$, $6\cdot 10^{-5} \le x \le 6\cdot 10^{-3}$ and
$p_T \le 200$~{\rm MeV}. Comparison of the proton and neutron data in the 
same kinematic domain shows that the production cross section for 
leading protons is larger than it is for leading 
neutrons. This result demonstrates that leading proton production cannot be 
entirely due to pion exchange. 

The LEPTO Monte Carlo program, which is based upon soft colour 
interactions and a string fragmentation model, describes the magnitude 
and the general shape of the neutron data with $z \ge 0.3$. 
It fails however to describe the rate of leading proton production and the
rise in the semi--inclusive structure function ${\mathrm{F_2^{LP(3)}}}$ as a 
function of $Q^2$. The RAPGAP Monte Carlo program, which simulates pion 
exchange, gives an acceptable description of the neutron data with 
$z \ge 0.7$ but does not explain the absolute rate of leading proton 
production. 

The proton and neutron data are equally well described by fits assuming
factorization in $x, Q^2$ and $z$ or $\beta, Q^2$ and $z$. The scaling
violations observed in the measured semi--inclusive structure functions
${\mathrm{F_2^{LP(3)}}}$ and ${\mathrm{F_2^{LN(3)}}}$ are similar in size
and shape and are close to those seen in the GRV parametrizations of the 
inclusive structure functions of the pion and the proton.

The neutron and proton data are reasonably well described by a Regge model 
of leading baryon production which considers the colour neutral exchanges of 
pions, pomerons and secondary reggeons. The semi--inclusive 
cross sections for leading neutrons with $0.7 \le z \le 0.9$ can be described 
entirely by $\pi^+$ exchange whereas the semi--inclusive cross sections for 
protons with $0.73 < z < 0.88$ require $\pi^0$ and $f_2$ 
exchange contributions. In our model, the contribution due to $f_2$ exchange
is approximately a factor of two greater than the contribution due to
$\pi^0$ exchange. The $\beta$ and $Q^2$ dependence of the leading
neutron data at $z = 0.7$ are consistent with the GRV leading order 
parton distributions for the pion.

\section*{Acknowledgements}

We are grateful to the HERA machine group whose outstanding
efforts have made and continue to make this experiment possible. We thank
the engineers and technicians for their work in constructing and now
maintaining the H1 detector, our funding agencies for financial support, the
DESY technical staff for continual assistance, and the DESY directorate for the
hospitality which they extend to the non--DESY members of the collaboration.
The forward proton spectrometer was supported by the INTAS93--43 project.

\newpage 

\begin{table}[b]
{\scriptsize
\begin{tabular}{|c|c|c|c|c|c|c|c|c|c|}
\hline
$x$      & $Q^2 \, {\rm [GeV^2]}$ & $z$ & ${\mathrm{F_2^{LP(3)}}} \pm \mathrm{stat} \pm \mathrm{sys}$ &
$x$      & $Q^2 \, {\rm [GeV^2]}$ & $z$ & ${\mathrm{F_2^{LP(3)}}} \pm \mathrm{stat} \pm \mathrm{sys}$ \\
\hline
 0.00010 &  2.5  & 0.732  & $ 0.0514 \pm  0.0086 \pm  0.0087 $  &  0.00104 &  7.5  & 0.732  & $ 0.0411 \pm  0.0125 \pm  0.0064 $ \\  
 0.00010 &  2.5  & 0.780  & $ 0.0590 \pm  0.0072 \pm  0.0079 $  &  0.00104 &  7.5  & 0.780  & $ 0.0859 \pm  0.0143 \pm  0.0100 $ \\  
 0.00010 &  2.5  & 0.829  & $ 0.0580 \pm  0.0071 \pm  0.0073 $  &  0.00104 &  7.5  & 0.829  & $ 0.0736 \pm  0.0131 \pm  0.0079 $ \\  
 0.00010 &  2.5  & 0.878  & $ 0.0499 \pm  0.0076 \pm  0.0112 $  &  0.00104 &  7.5  & 0.878  & $ 0.0631 \pm  0.0139 \pm  0.0136 $ \\  
\hline
 0.00033 &  2.5  & 0.732  & $ 0.0392 \pm  0.0058 \pm  0.0059 $  &  0.00329 &  7.5  & 0.732  & $ 0.0419 \pm  0.0148 \pm  0.0073 $ \\  
 0.00033 &  2.5  & 0.780  & $ 0.0550 \pm  0.0054 \pm  0.0060 $  &  0.00329 &  7.5  & 0.780  & $ 0.1126 \pm  0.0209 \pm  0.0157 $ \\  
 0.00033 &  2.5  & 0.829  & $ 0.0595 \pm  0.0057 \pm  0.0060 $  &  0.00329 &  7.5  & 0.829  & $ 0.0699 \pm  0.0167 \pm  0.0093 $ \\  
 0.00033 &  2.5  & 0.878  & $ 0.0528 \pm  0.0061 \pm  0.0112 $  &  0.00329 &  7.5  & 0.878  & $ 0.0832 \pm  0.0205 \pm  0.0191 $ \\  
\hline
 0.00104 &  2.5  & 0.732  & $ 0.0357 \pm  0.0076 \pm  0.0058 $  &  0.00104 & 13.3  & 0.732  & $ 0.1024 \pm  0.0230 \pm  0.0150 $ \\  
 0.00104 &  2.5  & 0.780  & $ 0.0354 \pm  0.0059 \pm  0.0045 $  &  0.00104 & 13.3  & 0.780  & $ 0.1168 \pm  0.0193 \pm  0.0121 $ \\  
 0.00104 &  2.5  & 0.829  & $ 0.0403 \pm  0.0064 \pm  0.0048 $  &  0.00104 & 13.3  & 0.829  & $ 0.1253 \pm  0.0200 \pm  0.0118 $ \\  
 0.00104 &  2.5  & 0.878  & $ 0.0461 \pm  0.0076 \pm  0.0102 $  &  0.00104 & 13.3  & 0.878  & $ 0.1114 \pm  0.0222 \pm  0.0233 $ \\  
\hline
 0.00033 &  4.4  & 0.732  & $ 0.0684 \pm  0.0123 \pm  0.0098 $  &  0.00329 & 13.3  & 0.732  & $ 0.0839 \pm  0.0211 \pm  0.0141 $ \\  
 0.00033 &  4.4  & 0.780  & $ 0.0807 \pm  0.0102 \pm  0.0080 $  &  0.00329 & 13.3  & 0.780  & $ 0.0715 \pm  0.0153 \pm  0.0094 $ \\  
 0.00033 &  4.4  & 0.829  & $ 0.1030 \pm  0.0117 \pm  0.0092 $  &  0.00329 & 13.3  & 0.829  & $ 0.0759 \pm  0.0164 \pm  0.0094 $ \\  
 0.00033 &  4.4  & 0.878  & $ 0.0887 \pm  0.0123 \pm  0.0183 $  &  0.00329 & 13.3  & 0.878  & $ 0.0500 \pm  0.0153 \pm  0.0112 $ \\  
\hline
 0.00104 &  4.4  & 0.732  & $ 0.0631 \pm  0.0121 \pm  0.0107 $  &  0.00104 & 28.6  & 0.732  & $ 0.0737 \pm  0.0279 \pm  0.0111 $ \\  
 0.00104 &  4.4  & 0.780  & $ 0.0619 \pm  0.0094 \pm  0.0083 $  &  0.00104 & 28.6  & 0.780  & $ 0.1030 \pm  0.0259 \pm  0.0112 $ \\  
 0.00104 &  4.4  & 0.829  & $ 0.0626 \pm  0.0091 \pm  0.0079 $  &  0.00104 & 28.6  & 0.829  & $ 0.0398 \pm  0.0163 \pm  0.0039 $ \\  
 0.00104 &  4.4  & 0.878  & $ 0.0572 \pm  0.0101 \pm  0.0129 $  &  0.00104 & 28.6  & 0.878  & $ 0.0840 \pm  0.0266 \pm  0.0178 $ \\  
\hline
 0.00033 &  7.5  & 0.732  & $ 0.0770 \pm  0.0189 \pm  0.0121 $  &  0.00329 & 28.6  & 0.732  & $ 0.1179 \pm  0.0307 \pm  0.0171 $ \\  
 0.00033 &  7.5  & 0.780  & $ 0.0652 \pm  0.0132 \pm  0.0077 $  &  0.00329 & 28.6  & 0.780  & $ 0.1047 \pm  0.0220 \pm  0.0106 $ \\  
 0.00033 &  7.5  & 0.829  & $ 0.0873 \pm  0.0151 \pm  0.0096 $  &  0.00329 & 28.6  & 0.829  & $ 0.0734 \pm  0.0184 \pm  0.0067 $ \\  
 0.00033 &  7.5  & 0.878  & $ 0.1080 \pm  0.0197 \pm  0.0234 $  &  0.00329 & 28.6  & 0.878  & $ 0.0784 \pm  0.0219 \pm  0.0163 $ \\  
\hline
\end{tabular}
\caption{The measured values of ${\mathrm{F_2^{LP(3)}}}(x,Q^2,z)$ for protons with $p_T~\le~200$~{\rm MeV}.
There is an additional normalization uncertainty of 5.6\% not included in the systematic error.}
\label{f2lpdata}
}
\end{table}

\newpage

\begin{table}[b]
{\scriptsize
\begin{tabular}{|c|c|c|c|c|c|c|c|c|c|}
\hline
$x$      & $Q^2 \, {\rm [GeV^2]}$ & $z$ & ${\mathrm{F_2^{LN(3)}}} \pm \mathrm{stat} \pm \mathrm{sys}$ &
$x$      & $Q^2 \, {\rm [GeV^2]}$ & $z$ & ${\mathrm{F_2^{LN(3)}}} \pm \mathrm{stat} \pm \mathrm{sys}$ \\
\hline                                                                                                      
 0.00010  & 2.5  & 0.3  &  $ 0.0829  \pm   0.0051  \pm   0.0219  $  &     0.00104  &  7.5  & 0.3  & $  0.0984  \pm   0.0085  \pm   0.0259  $    \\
 0.00010  & 2.5  & 0.5  &  $ 0.0470  \pm   0.0033  \pm   0.0165  $  &     0.00104  &  7.5  & 0.5  & $  0.0527  \pm   0.0054  \pm   0.0184  $    \\
 0.00010  & 2.5  & 0.7  &  $ 0.0396  \pm   0.0030  \pm   0.0068  $  &     0.00104  &  7.5  & 0.7  & $  0.0488  \pm   0.0047  \pm   0.0082  $    \\
 0.00010  & 2.5  & 0.9  &  $ 0.0125  \pm   0.0023  \pm   0.0073  $  &     0.00104  &  7.5  & 0.9  & $  0.0080  \pm   0.0029  \pm   0.0046  $    \\
\hline
 0.00033  & 2.5  & 0.3  &  $ 0.0673  \pm   0.0032  \pm   0.0176  $  &     0.00329  &  7.5  & 0.3  & $  0.0812  \pm   0.0079  \pm   0.0213  $    \\
 0.00033  & 2.5  & 0.5  &  $ 0.0378  \pm   0.0020  \pm   0.0132  $  &     0.00329  &  7.5  & 0.5  & $  0.0458  \pm   0.0051  \pm   0.0160  $    \\
 0.00033  & 2.5  & 0.7  &  $ 0.0315  \pm   0.0018  \pm   0.0052  $  &     0.00329  &  7.5  & 0.7  & $  0.0412  \pm   0.0043  \pm   0.0069  $    \\
 0.00033  & 2.5  & 0.9  &  $ 0.0066  \pm   0.0013  \pm   0.0038  $  &     0.00329  &  7.5  & 0.9  & $  0.0033  \pm   0.0028  \pm   0.0019  $    \\
\hline                                                                                                                                                      
 0.00104  & 2.5  & 0.3  &  $ 0.0582  \pm   0.0034  \pm   0.0152  $  &     0.00104  & 13.3  & 0.3  & $  0.1295  \pm   0.0121  \pm   0.0341  $    \\
 0.00104  & 2.5  & 0.5  &  $ 0.0296  \pm   0.0021  \pm   0.0103  $  &     0.00104  & 13.3  & 0.5  & $  0.0749  \pm   0.0081  \pm   0.0262  $    \\
 0.00104  & 2.5  & 0.7  &  $ 0.0257  \pm   0.0019  \pm   0.0043  $  &     0.00104  & 13.3  & 0.7  & $  0.0533  \pm   0.0073  \pm   0.0090  $    \\
 0.00104  & 2.5  & 0.9  &  $ 0.0059  \pm   0.0013  \pm   0.0034  $  &     0.00104  & 13.3  & 0.9  & $  0.0202  \pm   0.0063  \pm   0.0117  $    \\
\hline                                                                                                                                                      
 0.00033  & 4.4  & 0.3  &  $ 0.1150  \pm   0.0073  \pm   0.0302  $  &     0.00329  & 13.3  & 0.3  & $  0.0918  \pm   0.0082  \pm   0.0242  $    \\
 0.00033  & 4.4  & 0.5  &  $ 0.0664  \pm   0.0047  \pm   0.0232  $  &     0.00329  & 13.3  & 0.5  & $  0.0512  \pm   0.0054  \pm   0.0179  $    \\
 0.00033  & 4.4  & 0.7  &  $ 0.0526  \pm   0.0039  \pm   0.0088  $  &     0.00329  & 13.3  & 0.7  & $  0.0457  \pm   0.0048  \pm   0.0077  $    \\
 0.00033  & 4.4  & 0.9  &  $ 0.0091  \pm   0.0028  \pm   0.0053  $  &     0.00329  & 13.3  & 0.9  & $  0.0141  \pm   0.0037  \pm   0.0082  $    \\
\hline                                                                                                                                                      
 0.00104  & 4.4  & 0.3  &  $ 0.0952  \pm   0.0055  \pm   0.0251  $  &     0.00104  & 28.6  & 0.3  & $  0.1650  \pm   0.0224  \pm   0.0444  $    \\
 0.00104  & 4.4  & 0.5  &  $ 0.0450  \pm   0.0035  \pm   0.0157  $  &     0.00104  & 28.6  & 0.5  & $  0.0952  \pm   0.0140  \pm   0.0337  $    \\
 0.00104  & 4.4  & 0.7  &  $ 0.0414  \pm   0.0030  \pm   0.0070  $  &     0.00104  & 28.6  & 0.7  & $  0.0509  \pm   0.0137  \pm   0.0091  $    \\
 0.00104  & 4.4  & 0.9  &  $ 0.0101  \pm   0.0019  \pm   0.0059  $  &     0.00104  & 28.6  & 0.9  & $  0.0433  \pm   0.0127  \pm   0.0253  $    \\
\hline                                                                                                                                                      
 0.00033  & 7.5  & 0.3  &  $ 0.1228  \pm   0.0116  \pm   0.0327  $  &     0.00329  & 28.6  & 0.3  & $  0.1018  \pm   0.0129  \pm   0.0267  $    \\
 0.00033  & 7.5  & 0.5  &  $ 0.0752  \pm   0.0075  \pm   0.0265  $  &     0.00329  & 28.6  & 0.5  & $  0.0702  \pm   0.0086  \pm   0.0245  $    \\
 0.00033  & 7.5  & 0.7  &  $ 0.0442  \pm   0.0064  \pm   0.0077  $  &     0.00329  & 28.6  & 0.7  & $  0.0499  \pm   0.0077  \pm   0.0084  $    \\
 0.00033  & 7.5  & 0.9  &  $ 0.0220  \pm   0.0050  \pm   0.0128  $  &     0.00329  & 28.6  & 0.9  & $  0.0148  \pm   0.0071  \pm   0.0086  $    \\
\hline
\end{tabular}
\caption{The measured values of ${\mathrm{F_2^{LN(3)}}}(x,Q^2,z)$ for neutrons with $p_T~\le~200$~{\rm MeV}. 
There is an additional normalization uncertainty of 5.7\% not included in the
systematic error.}
\label{f2lndata}
}
\end{table}

\newpage 

\end{document}

%% file: h1auts.tex
\noindent
 C.~Adloff$^{34}$,                
 M.~Anderson$^{22}$,              
 V.~Andreev$^{25}$,               
 B.~Andrieu$^{28}$,               
 V.~Arkadov$^{35}$,               
 C.~Arndt$^{11}$,                 
 I.~Ayyaz$^{29}$,                 
 A.~Babaev$^{24}$,                
 J.~B\"ahr$^{35}$,                
 P.~Baranov$^{25}$
% ...

%% file: h1inst.tex
 $ ^1$ I. Physikalisches Institut der RWTH, Aachen, Germany$^a$ \\
 $ ^2$ III. Physikalisches Institut der RWTH, Aachen, Germany$^a$ \\
 $ ^3$ School of Physics and Space Research, University of Birmingham,
       Birmingham, UK$^b$\\
 $ ^4$ Inter-University Institute for High Energies ULB-VUB, Brussels;
       Universitaire Instelling Antwerpen, Wilrijk; Belgium$^c$ \\
 $ ^5$ Rutherford Appleton Laboratory, Chilton, Didcot, UK$^b$ \\
 $ ^6$ Institute for Nuclear Physics, Cracow, Poland$^d$  \\
 $ ^7$ Physics Department and IIRPA,
       University of California, Davis, California, USA$^e$ \\
 $ ^8$ Institut f\"ur Physik, Universit\"at Dortmund, Dortmund,
       Germany$^a$ \\
 $ ^9$ Joint Institute for Nuclear Research, Dubna, Russia \\
 $ ^{10}$ DSM/DAPNIA, CEA/Saclay, Gif-sur-Yvette, France \\
 $ ^{11}$ DESY, Hamburg, Germany$^a$ \\
 $ ^{12}$ II. Institut f\"ur Experimentalphysik, Universit\"at Hamburg,
          Hamburg, Germany$^a$  \\
 $ ^{13}$ Max-Planck-Institut f\"ur Kernphysik,
          Heidelberg, Germany$^a$ \\
 $ ^{14}$ Physikalisches Institut, Universit\"at Heidelberg,
          Heidelberg, Germany$^a$ \\
 $ ^{15}$ Institut f\"ur Hochenergiephysik, Universit\"at Heidelberg,
          Heidelberg, Germany$^a$ \\
 $ ^{16}$ Institut f\"ur experimentelle und angewandte Physik, 
          Universit\"at Kiel, Kiel, Germany$^a$ \\
 $ ^{17}$ Institute of Experimental Physics, Slovak Academy of
          Sciences, Ko\v{s}ice, Slovak Republic$^{f,j}$ \\
 $ ^{18}$ School of Physics and Chemistry, University of Lancaster,
          Lancaster, UK$^b$ \\
 $ ^{19}$ Department of Physics, University of Liverpool, Liverpool, UK$^b$ \\
 $ ^{20}$ Queen Mary and Westfield College, London, UK$^b$ \\
 $ ^{21}$ Physics Department, University of Lund, Lund, Sweden$^g$ \\
 $ ^{22}$ Department of Physics and Astronomy, 
          University of Manchester, Manchester, UK$^b$ \\
 $ ^{23}$ CPPM, Universit\'{e} d'Aix-Marseille~II,
          IN2P3-CNRS, Marseille, France \\
 $ ^{24}$ Institute for Theoretical and Experimental Physics,
          Moscow, Russia \\
 $ ^{25}$ Lebedev Physical Institute, Moscow, Russia$^{f,k}$ \\
 $ ^{26}$ Max-Planck-Institut f\"ur Physik, M\"unchen, Germany$^a$ \\
 $ ^{27}$ LAL, Universit\'{e} de Paris-Sud, IN2P3-CNRS, Orsay, France \\
 $ ^{28}$ LPNHE, \'{E}cole Polytechnique, IN2P3-CNRS, Palaiseau, France \\
 $ ^{29}$ LPNHE, Universit\'{e}s Paris VI and VII, IN2P3-CNRS,
          Paris, France \\
 $ ^{30}$ Institute of  Physics, Academy of Sciences of the
          Czech Republic, Praha, Czech Republic$^{f,h}$ \\
 $ ^{31}$ Nuclear Center, Charles University, Praha, Czech Republic$^{f,h}$ \\
 $ ^{32}$ INFN Roma~1 and Dipartimento di Fisica,
          Universit\`a Roma~3, Roma, Italy \\
 $ ^{33}$ Paul Scherrer Institut, Villigen, Switzerland \\
 $ ^{34}$ Fachbereich Physik, Bergische Universit\"at Gesamthochschule
          Wuppertal, Wuppertal, Germany$^a$ \\
 $ ^{35}$ DESY, Institut f\"ur Hochenergiephysik, Zeuthen, Germany$^a$ \\
 $ ^{36}$ Institut f\"ur Teilchenphysik, ETH, Z\"urich, Switzerland$^i$ \\
 $ ^{37}$ Physik-Institut der Universit\"at Z\"urich,
          Z\"urich, Switzerland$^i$ \\
\smallskip
 $ ^{38}$ Institut f\"ur Physik, Humboldt-Universit\"at,
          Berlin, Germany$^a$ \\
 $ ^{39}$ Rechenzentrum, Bergische Universit\"at Gesamthochschule
          Wuppertal, Wuppertal, Germany$^a$ \\
 $ ^{40}$ Vistor from Yerevan Physics Institute, Armenia \\
 $ ^{41}$ Foundation for Polish Science fellow \\
 $ ^{42}$ Institut f\"ur Experimentelle Kernphysik, Universit\"at Karlsruhe,
          Karlsruhe, Germany \\
 $ ^{43}$ Dept. F\'\i s. Ap. CINVESTAV, 
          M\'erida, Yucat\'an, M\'exico

 
\bigskip
\noindent
 $ ^a$ Supported by the Bundesministerium f\"ur Bildung, Wissenschaft,
        Forschung und Technologie, FRG,
        under contract numbers 7AC17P, 7AC47P, 7DO55P, 7HH17I, 7HH27P,
        7HD17P, 7HD27P, 7KI17I, 6MP17I and 7WT87P \\
 $ ^b$ Supported by the UK Particle Physics and Astronomy Research
       Council, and formerly by the UK Science and Engineering Research
       Council \\
 $ ^c$ Supported by FNRS-FWO, IISN-IIKW \\
 $ ^d$ Partially supported by the Polish State Committee for Scientific 
       Research, grant no. 115/E-343/SPUB/P03/002/97 and
       grant no. 2P03B~055~13 \\
 $ ^e$ Supported in part by US~DOE grant DE~F603~91ER40674 \\
 $ ^f$ Supported by the Deutsche Forschungsgemeinschaft \\
 $ ^g$ Supported by the Swedish Natural Science Research Council \\
 $ ^h$ Supported by GA~\v{C}R  grant no. 202/96/0214,
       GA~AV~\v{C}R  grant no. A1010821 and GA~UK  grant no. 177 \\
 $ ^i$ Supported by the Swiss National Science Foundation \\
 $ ^j$ Supported by VEGA SR grant no. 2/5167/98 \\
 $ ^k$ Supported by Russian Foundation for Basic Research 
       grant no. 96-02-00019 